\documentstyle[12pt]{article}
\renewcommand{\thefootnote}{\fnsymbol{footnote}}
\begin{document}
\begin{flushright}
Columbia preprint CU--TP--692
\end{flushright}
\vspace*{1cm}
\setcounter{footnote}{1}
\begin{center}
{\Large\bf Relativistic Hydrodynamics for Heavy--Ion Collisions}\\
{\Large\bf I. General Aspects and Expansion into
Vacuum\footnote{This work was supported by the Director, Office of Energy
Research, Division of Nuclear Physics of the Office of High Energy and Nuclear
Physics of the U.S. Department of Energy under Contract No.\
DE-FG-02-93ER-40764.}}
\\[1cm]
Dirk H.\ Rischke\footnote{Partially supported by the
Alexander von Humboldt--Stiftung under
the Feodor--Lynen program.} \\ ~~ \\
{\small Physics Department, Pupin
Physics Laboratories, Columbia University} \\
{\small 538 W 120th Street, New
York, NY 10027, U.S.A.} \\ ~~ \\
Stefan Bernard, Joachim A.\ Maruhn \\ ~~ \\
{\small Institut f\"ur Theoretische Physik der J.W.\ Goethe--Universit\"at} \\
{\small Robert--Mayer--Str.\ 10, D--60054 Frankfurt/M., Germany}
\\ ~~ \\ ~~ \\
{\large April 1995}
\\[1cm]
\end{center}
\begin{abstract}
We present algorithms to solve relativistic hydrodynamics
in 3+1--dimensional situations without apparent symmetry to simplify
the solution. In simulations of heavy--ion collisions, these
numerical schemes have to deal with the physical vacuum
and with equations of state with a first order phase transition
between hadron matter and a quark--gluon plasma. We investigate their
performance for the one--dimensional
expansion of baryon-free nuclear matter into the vacuum, which is an
analytically solvable
test problem that incorporates both the aspect of the vacuum as
well as that of a phase transition in the equation of state.
The dependence of the lifetime of the mixed phase on the
initial energy density is discussed.
\end{abstract}
\renewcommand{\thefootnote}{\arabic{footnote}}
\setcounter{footnote}{0}
\newpage
\section{Introduction}

Hydrodynamics represents (local) energy, momentum, and charge
conservation \cite{stoecker,strottman}. Because of its simplicity
it has found widespread application in studying the evolution of
nuclear matter in heavy--ion collisions. The physical input
is the nuclear matter equation of state (EoS).
Therefore, hydrodynamics allows to study nuclear matter properties
like the phase transition to the quark--gluon plasma (QGP)
\cite{QGP} in a simple, straightforward manner.

One of the first applications of hydrodynamics to heavy--ion
collisions was the study of multi--particle production by Fermi and
Landau \cite{landau}\footnote{Over the past years
the original Landau model and also simplified versions
(so-called ``fireball'' models) have been frequently applied to
explain heavy--ion collision data. To discuss these approaches in
greater detail is out of the scope of this paper.}.
Simple hydrodynamical models \cite{scheid} predicted the occurrence of
nuclear shock waves \cite{shock} in the early seventies.
Full 3+1--dimensional calculations for heavy--ion collisions
at BEVALAC energies ($E_{Lab}^{kin}= 0.1 - 2$ AGeV) were performed
about twenty years ago (see the very detailed review of
this topic in \cite{stoecker}, and refs.\ therein).
It was found that the compressional shock waves created in the collision
lead to collective flow phenomena like sideward deflection
of matter in the reaction plane (``side-splash'' and ``bounce-off'')
as well as azimuthal deflection out of the reaction plane (``squeeze-out'').

One of the main successes of the fluid-dynamical picture was the
confirmation of these collective flow effects by BEVALAC experiments
\cite{gutbrod}. More detailed investigations of the flow, however, have
revealed \cite{schuermann} that viscosity might be important to explain
these phenomena quantitatively. Also, systematic hydrodynamical
studies were done only in the non-relativistic limit.
Therefore, ideal and
dissipative relativistic hydrodynamical investigations
for BEVALAC energies are
mandatory. A comparison with recent excellent triple differential data
from the EOS--collaboration \cite{rai} would help
to understand in how far the hydrodynamical picture
is quantitatively applicable at these energies.

3+1--dimensional ideal relativistic hydrodynamics was applied to
study heavy--ion collisions at AGS ($E_{Lab}^{kin} \simeq 11$ AGeV)
and CERN--SPS energies ($E_{Lab}^{kin} \simeq 60 - 200$ AGeV)
\cite{rentzsch}. Since ideal hydrodynamics
assumes that matter is in local equilibrium at every instant,
colliding fluid elements are forced by
momentum conservation to instantaneously stop
and by energy conservation to convert all their kinetic energy
into thermal energy (compression and heating via shock waves).
However, the longitudinal rapidity loss in
individual nucleon--nucleon collisions is limited. Thus, immediate
complete stopping is not achieved in reality and, for higher beam
energies, it is no longer justified to treat the initial stage of the reaction
in an ideal hydrodynamical picture. Ideal hydrodynamics might nevertheless
be applicable in the expansion stage of the collision \cite{prakash},
where the conditions of local thermodynamical equilibrium
are more likely to be established. In order
to describe the initial stage, however, one has to extend the
hydrodynamical description to account for non-equilibrium effects.

While dissipative hydrodynamics in principle provides the theoretical
framework for the study of non-equilibrium effects, its
viability as a causal theory is still under debate \cite{strottman}.
Moreover, all known formulations which are rigorously derived
from kinetic theory rely on the assumption that non-equilibrium effects
can be treated as perturbations, or in other words,
that the momentum distribution of particles
deviates only slightly from its local equilibrium form.

This is not the case in the initial stage of heavy--ion collisions where
incident energies can be much larger than the Fermi--momentum of the nucleons,
and thus, the momentum distributions of target and projectile nucleons
are strongly displaced.
This has led to the development of the so-called multi--fluid-dynamical
approach to heavy--ion collisions \cite{amsden,katscher}.
Here the projectile and target nuclei are treated as ideal fluids,
but the collisional interaction between them is calculated via kinetic theory.
In the three--fluid model of Ref.\ \cite{katscher} the energy and
momentum loss in these collisions serve as source terms for a
third fluid which is also assumed to be in local equilibrium.
Thus, apart from its creation mechanism
(and from rescattering processes while it overlaps in space--time with the
first two fluids),
the dynamical evolution of this
fluid is similar to that in the Landau model.
For the AGS to CERN--SPS energy range this model was shown to
reproduce a variety of experimental observables, like meson rapidity
distributions as well as meson and photon transverse momentum spectra
\cite{adrian}. Of particular interest would be a comparison of
the hydrodynamical flow predicted by this model with the corresponding
experimental data at the AGS \cite{Yingchao}.

Finally, hydrodynamics was applied to the ultrarelativistic energy regime
(RHIC and LHC energies, $E_{CM} \geq 100$ AGeV) as well.
In this case, the nuclear stopping power is even for large nuclei
not sufficient\footnote{See, however, the results of \cite{schoenfeld}.}
to stop at least a few nucleons completely and
the nuclei will essentially pass through each other, leading
to a baryon-free region around midrapidity
$y \simeq 0$. However, individual parton collisions copiously generate
gluons which form a region of high energy density in that part of
phase space.
This picture is supported by event generators based on parton--parton
interactions \cite{PCM,HIJING} and has led to the proposal of the so-called
``hot--glue scenario'' \cite{shuryak2}.
After a (proper) time of order $1$ fm, matter is supposed to
establish local thermodynamical equilibrium and the subsequent dynamical
evolution can be described by hydrodynamics.

This picture was anticipated long ago by Bjorken \cite{bjorken} who suggested
a one--dimensional hydrodynamic scaling solution
(i.e., where the longitudinal velocity $v_z \equiv z/t$) for the evolution of
matter in the central region. Over the years this picture was
refined to include transverse motion \cite{baym} and the hadronization phase
transition \cite{friman,kataja,blaizot}. Due to its simplicity, it is
sometimes even used to explain experimental
observables for heavy--ion collisions at the SPS \cite{dinesh},
although at these energies neither is there a baryon-free region
at mid-rapidity \cite{SPS} nor is
the longitudinal motion likely to be of the scaling type.

Recent investigations with the HIJING event generator \cite{mgdhr}
have revealed that for RHIC energies
the initial conditions for the hydrodynamic stage
of the system's evolution might not be homogeneous on surfaces of
constant proper time, as in Bjorken's picture \cite{bjorken},
but subject to strong fluctuations\footnote{Actually, this possibility
was already pointed out by Bjorken \cite{bjorken}.}. As was argued
in \cite{mgdhr}, hydrodynamics might still be applicable to study the
subsequent evolution, but the assumptions of scaling for the longitudinal
motion and cylindrical symmetry for the transverse motion are
certainly no longer valid.

All these more recent developments call for schemes to solve hydrodynamics,
which are applicable in the general case, i.e.,
without obvious space--time symmetries that help to simplify the solution.
Such schemes are necessarily of numerical nature.
In a previous work \cite{schneider}, several numerical algorithms
for relativistic hydrodynamics were presented and shown to perform well
for analytically solvable test cases like the Riemann problem for
matter with an ideal gas EoS. These investigations present a necessary first
step in the development of a hydrodynamical scheme for
simulations of heavy--ion collisions.
Following this line we consider in this work the ability
of the algorithms to cope with two problems specifically occurring
in simulations of heavy--ion collisions.
The first is the presence of vacuum and the second the fact
that a phase transition in the EoS changes the character of the
hydrodynamical solution.

Vacuum in relativistic hydrodynamics poses the following problem:
matter can at most propagate with the velocity of light into the vacuum, i.e.,
during a time $\Delta t$ matter travels at most a distance $\Delta t$
into the vacuum\footnote{We use natural units $\hbar=c=k_B=1$.}.
The equations of relativistic hydrodynamics respect
this causality requirement since they are hyperbolic.
On the other hand, any algorithm solving the finite difference form of a
hyperbolic differential equation has to fulfill the
Courant--Friedrichs--Lewy (CFL) criterion \cite{CFL}, namely
that the ratio of time step width to spatial grid spacing is less than unity,
$\Delta t / \Delta x \equiv \lambda < 1$. Causal matter
transport covers at most a distance
$\Delta t \equiv \lambda \Delta x < \Delta x$ in one time step.
However, the algorithms {\em average\/} transported quantities
over a cell after each time step. Therefore, even if the algorithm
transports a certain amount of matter causally into the vacuum {\em
during\/} a time step, it will be evenly distributed over a spatial distance
$\Delta x > \lambda \Delta x$ at the {\em end\/} of the time step.
Thus, a part of matter is acausally propagated a distance
$ (1-\lambda) \Delta x$ into the space-like light cone. We will
henceforth call this purely numerical phenomenon {\em prediffusion}.
Obviously, the larger $\lambda$, the smaller this prediffusion. However,
some Flux Corrected Transport (FCT) algorithms, like the SHASTA
\cite{SHASTA}, require $\lambda < 1/2$ and thus might
lead to a considerable amount of prediffusion. We will consider the
prediffusion of this algorithm and also of the relativistic
Harten--Lax--van Leer--Einfeldt (HLLE) scheme \cite{schneider},
a Godunov-type algorithm \cite{holt}, for the one--dimensional expansion of
matter into vacuum.

The second problem is related to one of the main
advantages of the hydrodynamical approach over
microscopic models, namely the ability to dynamically study
the phase transition to the QGP in a relatively straightforward way:
this phase transition modifies to first approximation (neglecting
non-equilibrium phenomena such as supercooling and bubble formation
\cite{bubble}) only the EoS which is given input to
the hydrodynamical equations of motion, while a microscopic model
would have to implement a microscopic mechanism for hadronization
and deconfinement \cite{werner}.

At present, the order of the phase transition is not clear \cite{pt}.
In the absence of more fundamental theoretical information, especially for
baryon-rich matter, one commonly resorts to the MIT bag EoS
\cite{MIT} for the QGP which is matched to a state-of-the-art hadron matter
EoS \cite{hadron} via Gibbs conditions of phase equilibrium
(see also \cite{test2}). Thus, the transition is first order by construction.

However, matter with a first order phase transition may exhibit
thermodynamically anomalous behaviour in a certain range
of independent thermodynamic variables. Such behaviour is
signalled by a change of sign of the quantity
\begin{equation} \label{Sigma}
\Sigma \equiv \left. \frac{\partial^2 p}{\partial \epsilon^2} \right|_{\sigma}
 + 2\, c_s^2~\frac{1-c_s^2}{\epsilon +p}~,
\end{equation}
where $c_s^2 \equiv \partial p/ \partial \epsilon |_{\sigma}$ is the
velocity of sound, $\epsilon, p,$ and $\sigma$ are the energy density,
pressure, and specific entropy, respectively \cite{bugaev}.
For thermodynamically normal (TN) matter, $\Sigma >0$, for
so-called thermodynamically anomalous (TA) matter $\Sigma < 0$
\cite{bugaev}.
For the expansion of TN matter a simple rarefaction wave is the stable
hydrodynamical solution and for TA matter it is a rarefaction shock wave
(see Section 4).
On the other hand, for the compression of TN matter a compressional shock wave
forms the stable hydrodynamic solution while for TA matter
this is a compressional simple wave (the compression of
TA matter is investigated in a sequel to this paper \cite{test2}).

An example for matter with TA behaviour is (net) baryon-free
nuclear matter with an EoS of a
(massless) pion gas at low temperatures and a QGP
(described by the MIT bag EoS) at high temperatures
with a first order phase transition between these phases
at the temperature $T_c$.
While the pure QGP and pionic phase are TN, the first order
phase transition between them gives rise to a mixed phase
where $\Sigma$ changes its sign
and, thus, where the EoS becomes TA\footnote{In a strict sense,
$\Sigma$ only becomes zero for this EoS. For the expansion problem
studied here, this nevertheless leads
to a discontinuity in the hydrodynamic expansion, see Section 4 for details.}.
For the one--dimensional
expansion of matter
into the vacuum we investigate how well the numerical algorithms
are able to accomodate to this change in the thermodynamic behaviour
and reproduce the correct pattern of rarefaction, i.e., a
simple rarefaction wave in the TN region and a rarefaction
shock wave in the TA region of the EoS.

Our considerations focus first on the one--dimensional
expansion of semi-infinite matter, because in this case
an analytical solution exists
both for TN {\em and\/} TA matter, against which the results produced by
the numerical algorithms can be compared.
We then study the one--dimensional
expansion of matter extending over a spatially
finite region such as occurs in heavy--ion collisions and
discuss the lifetime of the mixed phase as a function of the initial
energy density.

This paper is organized as follows. In Section 2 we review
the equations of ideal relativistic hydrodynamics. Since we want
to keep this presentation as self-contained as possible, to
guarantee reproducibility of our results, and to facilitate
and encourage the use of the two algorithms considered here,
Section 3 gives an exact description of the SHASTA and the relativistic HLLE
as employed here.
In Section 4 we construct the analytical solutions for the one--dimensional
expansion
of both TN and TA semi-infinite matter into the vacuum. Section 5
contains our results on the reproduction of these results by
the numerical algorithms. In Section 6 we discuss
finite expanding systems and Section 7 closes this work with a
summary of our results. An Appendix presents modifications
of the SHASTA and their performance for some of the test
problems of Section 5.

\section{Ideal relativistic hydrodynamics}
The relativistic hydrodynamical equations represent
(local) energy--momentum conservation
\begin{equation} \label{dt}
\partial_{\mu} T^{\mu \nu} =0
\end{equation}
and (local) charge conservation
\begin{equation} \label{dn}
\partial_{\mu} N^{\mu} = 0~.
\end{equation}
For heavy--ion collisions, the conserved charge is e.g.\ the (net)
baryon number, the (net) strangeness etc. and there is an equation
of the type (\ref{dn}) for each conserved charge.
Provided that matter is in local thermodynamical equilibrium, the
energy--momentum tensor $T^{\mu \nu}$ and the charge current $N^{\mu}$
assume ideal fluid form \cite{LL6}, i.e.,
\begin{eqnarray} \label{t}
T^{\mu \nu} & = & (\epsilon +p) u^{\mu} u^{\nu} - p g^{\mu \nu}~,\\
N^{\mu} & = & n u^{\mu}~, \label{n}
\end{eqnarray}
where $\epsilon, p,$ and $n$ are energy density, pressure, and charge
density in the local rest frame of the fluid, $u^{\mu} \equiv \gamma
(1,{\bf v})$ is the fluid 4--velocity ($\gamma \equiv (1-{\bf v}^2)^{-1/2}$,
${\bf v}$ is the fluid 3--velocity), and $g^{\mu \nu} = {\rm diag}(+,-,-,-)$
is the metric tensor.
The equations of ideal fluid-dynamics are closed by specifying an EoS
for the matter under consideration in the form $p=p(\epsilon,n)$.

For the numerical implementation discussed in the next section it
is convenient to write the energy density in the
calculational frame as $T^{00} \equiv E$, the momentum density
as $T^{0i} \equiv M^i$, and the charge density as
$N^0\equiv R$, respectively. The equations of motion (\ref{dt},
\ref{dn}) then take the form
\begin{eqnarray} \label{dE}
\partial_t\, E + \mbox{\boldmath $\nabla$} \cdot (E \, {\bf v}) & = & -
\mbox{\boldmath $\nabla$} \cdot (p\, {\bf v})~, \\
\partial_t\, {\bf M} + \mbox{\boldmath $\nabla$} \cdot ({\bf M \, v}) & = &
-\mbox{\boldmath $\nabla$} p~,\\
\partial_t\, R + \mbox{\boldmath $\nabla$} \cdot (R \, {\bf v}) & = & 0~.
\label{dR}
\end{eqnarray}

\section{Numerical algorithms to solve relativistic hydrodynamics}

In numerical applications, one solves finite difference versions of
eqs.\ (\ref{dE}--\ref{dR}). The algorithms discussed here are Eulerian, i.e.,
the quantities $E, {\bf M},$ and $R$ are discretized on a fixed (and in
our case, Euclidean) grid in the calculational frame (which is conveniently
chosen as the global rest frame of the system under consideration).
They are also explicit, i.e., the calculation of quantities at the
(discrete) time step $n+1$ involves only quantities at previous time steps.

The first step is to treat the 3--divergence operator in
eqs.\ (\ref{dE}--\ref{dR})
via the method of time-step splitting (operator splitting), i.e.,
for instance eq.\ (\ref{dR}) is actually solved by sequentially
solving $\partial_t R + \partial_i (R\, v^i) = 0,\, i=x,y,z$. (No
summation over $i$ is implied. It is
advantageous to use permutations of the sequence $(x,y,z)$ in each
new time step.)
Therefore, the numerical algorithm is only required to solve equations
of the type
\begin{equation} \label{du}
\partial_t\, U + \partial_x (Uv +f) = 0~.
\end{equation}

\subsection{The SHASTA algorithm}
We define for further purpose (lower indices denote cell numbers,
upper indices the time step)
\begin{equation} \label{del}
\Delta_j \equiv U_{j+1}^n - U_j^n~.
\end{equation}
For the SHASTA \cite{SHASTA},
one first computes so-called transported and diffused quantities
\begin{equation} \label{Utd}
\tilde{U}_j= \frac{1}{2}\, Q_{+}^2\, \Delta_j
           - \frac{1}{2}\, Q_{-}^2\, \Delta_{j-1}
           + (Q_{+} + Q_{-})\, U_j^n + \lambda~\Delta f~,
\end{equation}
where
\begin{equation}
Q_{\pm} \equiv \frac{1/2 \mp \epsilon_j}{1 \pm (\epsilon_{j \pm 1} -
\epsilon_j)}~~,~~~\epsilon_j \equiv v_j^{n+1/2} \lambda~.
\end{equation}
Eq.\ (\ref{Utd}) results from
a simple graphical picture of mass transport over a fixed grid and
yields therefore a conservative transport scheme \cite{SHASTA}.
The source term $\Delta f$ may depend on the values of $f$
in cell $j$ and the neighbouring cells, like in the original explicit SHASTA
proposed by Boris and Book \cite{SHASTA},
\begin{equation} \label{orig}
\Delta f = - Q_{+} (f_{j+1}^{n+1/2} -f_j^{n+1/2}) -
Q_{-} (f_j^{n+1/2} - f_{j-1}^{n+1/2})~,
\end{equation}
however, we found that computing the source term with the much simpler
prescription
\begin{equation} \label{simsource}
\Delta f = - \frac{1}{2}~(f_{j+1}^{n+1/2} - f_{j-1}^{n+1/2})
\end{equation}
yields results superior to (\ref{orig}) (see Section 5 and Appendix).
The source terms and the velocities entering the
$\epsilon_j$ are computed at half time steps $n+1/2$ (half-step method)
to ensure second order accuracy in time.
Although much less time-consuming, simple first order full-step methods (which
take values at full time steps $n$ only) yield overshoots at shock fronts
(e.g.\ as occurring in the one--dimensional shock model studied in
\cite{schneider}) and should therefore not be used.

After transporting the quantity $U$ one has to remove the numerical
diffusion inherent in the transport scheme (\ref{Utd}).
Let us define the quantity
\begin{equation} \label{del2}
\tilde{\Delta}_j \equiv \tilde{U}_{j+1} - \tilde{U}_j~.
\end{equation}
The usual estimate \cite{SHASTA} for the diffusion in (\ref{Utd})
leads to the {\em antidiffusion\/} fluxes
\begin{equation} \label{Aex}
A_j^{ex} = \frac{1}{8}\, \tilde{\Delta}_j~,
\end{equation}
which are subtracted from (\ref{Utd}) to yield the final time-advanced
quantities $U_j^{n+1}$ (cf.\ eq.\ (\ref{Unp1}) below).
The superscript indicates ``explicit'', since this antidiffusion was used
in the original explicit SHASTA \cite{SHASTA}. However, we found the
antidiffusion of the so-called ``phoenical'' SHASTA \cite{SHASTA2}
to yield better results (see Section 5 and Appendix). For this version,
the antidiffusion flux reads
\begin{equation} \label{Aph}
A_j^{ph} = \frac{1}{8}\, \left(\tilde{\Delta}_j - \frac{1}{8}
\left[ \Delta_{j+1} - 2 \Delta_j + \Delta_{j-1} \right] \right)~.
\end{equation}
Of course, both for the explicit and the phoenical SHASTA,
the antidiffusion step should not create new maxima or minima.
This is taken into account by replacing the naive antidiffusion fluxes
(\ref{Aex}, \ref{Aph}) by the ``flux corrected'' antidiffusion fluxes
\begin{equation} \label{FCT}
\tilde{A}_j = \sigma \cdot {\rm max} \left\{ 0, {\rm min} \left[\sigma
\tilde{\Delta}_{j+1} , |A_j|, \sigma \tilde{\Delta}_{j-1} \right] \right\}~,
\end{equation}
$\sigma \equiv {\rm sgn}\, A_j$. The final time-advanced quantities read
\begin{equation} \label{Unp1}
U_j^{n+1} = \tilde{U}_j - \tilde{A}_j + \tilde{A}_{j-1}~.
\end{equation}
The variables $U^{n+1}_j$ should fulfill
the relativistic constraint $E \geq |{\bf M}|$. A simple way to ensure this
is discussed at the end of the next subsection.

\subsection{The relativistic HLLE algorithm}

The relativistic HLLE algorithm was originally proposed in \cite{schneider}
and is a Godunov-type algorithm \cite{holt}.
A so-called {\em Godunov algorithm\/} \cite{holt} makes use of the fact
that hydrodynamic fields are piecewise
constant inside each cell and discontinuous at the
cell boundaries after each time step.
The {\em exact\/} time evolution for this particular
kind of initial condition is given by the decay of all these discontinuities.
The decay of a discontinuity is the so-called {\em Riemann problem\/} and is
in principle analytically solvable (see, for instance, Ref.\ \cite{schneider}
and references therein).
A Godunov algorithm simply employs this solution of the Riemann problem
for {\em each\/} cell boundary to solve the hydrodynamical equations over
the {\em whole\/} calculational grid (see, for instance, \cite{blaizot}).
In turn, {\em Godunov-type\/} algorithms \cite{holt,marti}
do not use the exact solution of
the Riemann problem at a cell boundary but approximate it
in a manner consistent with the integral
conservation laws.

To be more specific, we consider the cell boundary at $x=0$ and the initial
distribution of the field $U$ at time $t=0$,
\begin{equation}
U (x,0) \equiv \left\{ \begin{array}{ll}
                    U_l & ,~~ x <0 \\
                    U_r & ,~~  x \geq 0~,
                       \end{array} \right.
\end{equation}
with $U_l \neq U_r$. For times $t > 0$ the discontinuity at $x=0$
will decay and produce a distribution
\begin{equation} \label{U}
U (x,t) \equiv \left\{ \begin{array}{ll}
                    U_l & ,~~ x < b_l t \\
                    U_{lr}(x) & ,~~b_l t \leq x < b_r t \\
                    U_r & ,~~  x \geq b_r t~.
                       \end{array} \right.
\end{equation}
$b_l <0 $ and $b_r>0$ are the so-called signal velocities
with which the decay proceeds to the left or right, respectively.
It is obvious that one should consider times
$t \leq \Delta x /|b_{l,r}|$, since otherwise
the decay of the discontinuity reaches the next cell boundary.
If the algorithm fulfills the CFL criterion $\lambda=
\Delta t / \Delta x <1$ this is automatically avoided
(since the signal velocities are causal).

In the exact solution to the Riemann problem, $U_{lr}(x)$ is a
nontrivial function of $x$ (involving a simple rarefaction wave, a
contact discontinuity and a shock wave, separated by
regions of constant flow, see e.g.\ \cite{schneider}). The
relativistic HLLE algorithm approximates this function by a constant, which
is determined by inserting (\ref{U}) into
eq.\ (\ref{du}) and integrating over
a fixed interval $[x_{min},x_{max}],\, x_{min} < b_l t, \, x_{max} > b_r t$:
\begin{equation} \label{Ulr}
U_{lr} \equiv \frac{b_r U_r - b_l U_l - F(U_r) + F(U_l)}{b_r - b_l}~,
\end{equation}
where $F(U) \equiv Uv + f$.
Integrating eq.\ (\ref{du}) over the (fixed) interval
$[x_{min},0]$ or $[0,x_{max}]$
and using (\ref{Ulr}) yields a value for $F(U_{lr})$:
\begin{equation} \label{Flr}
F(U_{lr}) = \frac{b_r F(U_l) - b_l F(U_r) + b_l b_r (U_r -U_l)}{b_r - b_l}~.
\end{equation}
To determine time-advanced values $U_j^{n+1}$
for each cell in the relativistic HLLE algorithm, one first
writes (\ref{du}) in the finite difference form
\begin{equation} \label{Ujnp1}
U_j^{n+1} = U_j^n - \lambda \left( G_{j+1/2} - G_{j-1/2}\right)~.
\end{equation}
Taking $G_{j+1/2} \equiv F(U_{j+1/2}^n)$, where the latter is
given by eq.\ (\ref{Flr}), one obtains a HLLE scheme with first
order accuracy in time. To achieve second order accuracy (as
for the SHASTA), half-step updated values for the numerical fluxes
$G$ are determined as follows \cite{schneider}.
First, cell interface values at time step $n$ are obtained via
replacing the piecewise constant distribution of the $U_j^n$
by a piecewise linear one,
\begin{equation} \label{Upm}
U^n_{j\pm} \equiv U^n_j \pm \frac{1}{2}\, S(\Delta_{j-1},
\Delta_j)~,
\end{equation}
where $\Delta_j$ is defined in eq.\ (\ref{del}) and
\begin{equation} \label{S}
S (\Delta_{j-1},\Delta_j) = \left\{ \begin{array}{lll}
 \Delta_{j-1} & {\rm for}~~|\Delta_{j-1}| \leq |\Delta_j|
&,~\Delta_{j-1} \cdot \Delta_j >0 \\
 \Delta_j & {\rm for}~~|\Delta_{j-1}| > |\Delta_j|
&,~\Delta_{j-1} \cdot  \Delta_j >0 \\
 0 & {\rm otherwise} &.
                                        \end{array} \right.
\end{equation}
When applying (\ref{Upm}) one has also to take care of the fact that
the interface values must not violate the relativistic constraint $E \geq
|{\bf M}|$.
Then, half-step updated values $U_{j \pm}^{n+1/2}$ are estimated by
a simple difference form of (\ref{du}),
\begin{equation}
U_{j \pm}^{n+1/2} = U_{j \pm}^n - \frac{\lambda}{2} \left( F(U_{j+}^n)
- F(U_{j-}^n) \right)~.
\end{equation}
Again, these values should respect $ E \geq |{\bf M}|$.
Inserting these values on the right-hand side of
(\ref{Flr}) we obtain half-step updated values for $G_{j+1/2}$,
\begin{equation}
G_{j+1/2} = \frac{b_{j+1/2}^r~F\left( U_{j+}^{n+1/2}\right) - b_{j+1/2}^l~
F\left( U_{(j+1)-}^{n+1/2}\right) + b_{j+1/2}^l~b_{j+1/2}^r~\left(
U_{(j+1)-}^{n+1/2} - U_{j+}^{n+1/2}\right)}{b_{j+1/2}^r - b_{j+1/2}^l}~.
\end{equation}
The transport problem (\ref{Ujnp1}) is solved if we specify
the signal velocities $b^r_{j+1/2}$ and $b^l_{j+1/2}$ at
the cell interface $j+1/2$. We note that if the slope function
(\ref{S}) is globally zero, this second order scheme is reduced to
first order accuracy in time.

The physical velocities for signals travelling to the
right/left is the relativistic addition/subtraction of matter velocity
and velocity of sound, $(v \pm c_s) /(1 \pm v c_s)$.
It was shown in \cite{schneider} that a convenient choice
for the signal velocities is
\begin{eqnarray} \label{signal1}
b_{j+1/2}^r & = & {\rm max} \left\{ 0,~\frac{\bar{v} + \bar{c}_s}{
1 + \bar{v} \bar{c}_s },~\frac{v_{(j+1)-}^{n+1/2} + c_{s,(j+1)-}^{n+1/2} }{
1 + v_{(j+1)-}^{n+1/2} c_{s,(j+1)-}^{n+1/2}} \right\}~,\\
b_{j+1/2}^l & = & {\rm min} \left\{ 0,~\frac{\bar{v} - \bar{c}_s}{
1 - \bar{v} \bar{c}_s },~\frac{v_{j+}^{n+1/2} - c_{s,j+}^{n+1/2}}{
1 - v_{j+}^{n+1/2} c_{s,j+}^{n+1/2}} \right\}~. \label{signal2}
\end{eqnarray}
The max--(min--)function accounts for the proper sign of $b_r$ ($b_l$).
Here
\begin{eqnarray} \label{av1}
\bar{v} & \equiv & \frac{\sqrt{E_{(j+1)-}^{n+1/2}}~v_{(j+1)-}^{n+1/2}
                       + \sqrt{E_{j+}^{n+1/2}}~v_{j+}^{n+1/2}}{
                     \sqrt{E_{(j+1)-}^{n+1/2}} + \sqrt{E_{j+}^{n+1/2}} }~, \\
\bar{c}_s^2 & \equiv & \frac{\sqrt{E_{(j+1)-}^{n+1/2}}~\left(
  c_{s,(j+1)-}^{n+1/2}\right)^2   + \sqrt{E_{j+}^{n+1/2}}~\left(
  c_{j+}^{n+1/2} \right)^2}{\sqrt{E_{(j+1)-}^{n+1/2}} + \sqrt{E_{j+}^{n+1/2}} }
 \label{av2}\\
  & + & \eta~\frac{ \sqrt{E_{(j+1)-}^{n+1/2}E_{j+}^{n+1/2}} }{ \left(
  \sqrt{E_{(j+1)-}^{n+1/2}} + \sqrt{E_{j+}^{n+1/2}} \right)^2}
 \left(v_{(j+1)-}^{n+1/2} - v_{j+}^{n+1/2} \right)^2~. \nonumber
\end{eqnarray}
In contrast to \cite{schneider} we use $E$ instead of $R$ in the Roe--averages
(\ref{av1}, \ref{av2}), since for (net) baryon-free flow $R \equiv 0$.
For all problems studied $\eta = 0.5$ in equation (\ref{av2}) proves to
be a safe choice.
In regions of vacuum, where the denominator in equations
(\ref{av1}, \ref{av2}) vanishes, we use the simple estimate
$b^r_{j+1/2} = 1$ and $b^l_{j+1/2}=-1$ which is robust for
all cases considered.

This completes the description of the relativistic HLLE algorithm. Also
here, the final variables $U^{n+1}_j$ must obey $E \geq |{\bf M}|$.
We found that restricting $|{\bf M}|$ to $E$
for cells violating this constraint provides the simplest way to
ensure this. For all problems studied here, the effect on the global
conservation of energy and momentum is negligible.

\subsection{Transformation between CM and local rest frame}

Both algorithms discussed so far require the values of pressure and
velocity at various points in the transport algorithm. Since
$p=p(\epsilon, n)$, the set of local rest frame
variables $\epsilon,{\bf v},n$ has to be determined from
the calculational frame variables $ E,{\bf M},R$.
According to the definition of the latter, this requires in general
an inversion of the five equations
\begin{eqnarray} \label{E}
E & = & (\epsilon + p) \gamma^2 -p ~, \\
{\bf M} & = & (\epsilon + p) \gamma^2~{\bf v}~,\label{M} \\
R & = & n \gamma~. \label{R}
\end{eqnarray}
Nevertheless, it is not necessary to perform a five--dimensional
root search \cite{kaempfer}. First note that ${\bf M}$ and ${\bf v}$ point
into the same direction. Let $|{\bf M}| \equiv M, |{\bf v}| \equiv v$.
Now observe that $M \equiv (E+p)v$ and also
\begin{equation} \label{en}
\epsilon = E-Mv~,~~n=R(1-v^2)^{1/2}~.
\end{equation}
Thus,
\begin{equation}
v = \frac{M}{E+p\left(E-Mv,R(1-v^2)^{1/2}\right)}~.
\end{equation}
For given calculational frame quantities $E,M,R$, this is
a simple fixed--point equation for $v$ and can be solved
iteratively. Once $v$ is known, $\epsilon$ and $n$ can be inferred from
(\ref{en}) and $p$ from the EoS. The different components of ${\bf v}$
can be obtained from the collinearity of ${\bf M}$ and ${\bf v}$.
For a simple EoS, one might even find an analytical solution for $v$
\cite{schneider}. However, the advantage of the method described here
is that it is independent of the EoS
(as long as $p\geq 0$, so that $E\geq M$ ensures causal matter flow $v\leq 1$)
and thus (almost) universally applicable.

\section{The one--dimensional expansion into vacuum -- analytical solution}

In this section we discuss the one--dimensional expansion of
semi-infinite, (net) baryon-free TN and TA matter into the vacuum.
The initial condition is
\begin{eqnarray} \label{in1}
\epsilon (x,0) & = & \left\{ \begin{array}{ll}
                          \epsilon_0 &,~~x\leq 0\\
                          0 &,~~x > 0~,
                            \end{array} \right. \\
v(x,0) & = & \left\{ \begin{array}{ll}
                          0 &,~~x\leq 0\\
                          1 &,~~x > 0~.
                            \end{array} \right. \label{in2}
\end{eqnarray}
Other hydrodynamic variables follow from the EoS and relations
(\ref{E}--\ref{R}). The choice $v=1$
in the vacuum is purely conventional,
but it guarantees a continuous hydrodynamic solution at
the boundary to the vacuum, since in the limit of infinite dilution
the velocity of matter approaches unity\footnote{In multi--dimensional
applications, due to the isotropy of the vacuum it is not possible
to assign it a directed, finite
velocity. The only possible choice is then $v=0$.}.

\subsection{Relation between thermodynamic properties and type
of hydrodynamical solution}

Let us clarify the relationship between the thermodynamic properties
of matter, i.e., whether it is TN or TA, and the type of hydrodynamical
solution. Consider one--dimensional hydrodynamic flow.
Then one can show \cite{courant} that regions of constant flow can only
be separated by a simple wave (continuous solution of the
hydrodynamic equations), or by a shock or a contact discontinuity
(discontinuous solutions of the hydrodynamic equations).
This means that between the two regions of constant flow in
(\ref{in1}, \ref{in2})
either the initial discontinuity will be preserved, or a
single simple wave will form to connect these regions,
or a more complicated flow pattern occurs, involving
a sequence of simple waves, discontinuities, and constant flow.
Unless the pressure $p(\epsilon)$ vanishes for all $\epsilon$, and matter
does not expand at all due to the absence of a driving force, the first
case can be ruled out.

Now consider a simple wave, i.e., continuous hydrodynamic flow.
This flow respects entropy conservation \cite{LL6}.
For isentropic flow, the $t-$ and $x-$component of the equation of motion
(\ref{dt}) can be combined to yield
\begin{equation}
\left(\frac{\partial}{\partial t} + \frac{ v \pm c_s }{ 1 \pm v c_s}
\frac{\partial}{\partial x} \right) {\cal R}_{\pm} = 0~,
\end{equation}
where the so-called {\em Riemann invariants\/} are
\begin{equation} \label{Rpm}
{\cal R}_{\pm} \equiv  y - y_0 \pm \int_{\epsilon_0}^\epsilon \frac{ c_s \,
{\rm d}\epsilon' }{\epsilon' + p(\epsilon')}~.
\end{equation}
Here $y={\rm Artanh}\, v$ is the fluid rapidity.
Obviously, the Riemann invariants are constant,
\begin{equation}
\left. \frac{{\rm d} {\cal R}_{\pm}}{{\rm d} t} \right|_{{\cal C}_{\pm}}=0~,
\label{ch2}
\end{equation}
along the so-called {\em characteristic curves\/} ${\cal C}_{\pm}(x,t)$,
the positions $x_{\pm}(t)$ of which are determined by integrating
\begin{equation} \label{ch1}
w_{\pm} \equiv \frac{{\rm d}x_{\pm}}{{\rm d}t} = \frac{ v \pm c_s }{ 1 \pm
v c_s}~.
\end{equation}
The characteristics are the world-lines of sonic disturbances
on the hydrodynamic flow pattern. For simple waves moving to the right
(i.e., $v >0$),
one can prove \cite{courant} that ${\cal R}_+ = const.$, and it
suffices to consider the ${\cal C}_-$--characteristic. For simple
waves moving to the left ($v<0$), ${\cal R}_-=const.$, and only
the ${\cal C}_+$--characteristics have to be considered.

Consider now simple waves moving to the right (left) and
a bundle of the corresponding non-trivial characteristics (i.e.,
the ${\cal C}_{\mp}$--characteristics) in the $x-t-$plane.
For constant time $t$, the change with $x$ of the
inverse slope (\ref{ch1}) of these characteristics
in the $x-t-$plane is
\begin{equation}
\left. \frac{\partial w_{\mp}}{\partial x} \right|_t \equiv
w_{\mp}' = \frac{v' (1- c_s^2) \mp c_s' (1-v^2)}{(1 \mp v c_s)^2}~.
\end{equation}
For a simple wave moving to the right (left), we deduce from
${\cal R}_{\pm}=const.$
\begin{equation}
{\rm d} v = \mp (1-v^2)\, \frac{c_s}{\epsilon +p}\, {\rm d}\epsilon~,
\end{equation}
and with
\begin{equation}
{\rm d} c_s = \frac{1}{2c_s} \left. \frac{\partial^2 p}{\partial \epsilon^2}
\right|_{\sigma} {\rm d} \epsilon
\end{equation}
and eqs.\ (\ref{Sigma}, \ref{ch1}) we obtain
\begin{equation}
w_{\mp}' = \mp\, \frac{1- w_{\mp}^2}{2 \, c_s \, (1-c_s^2)}~\Sigma~\epsilon'~.
\end{equation}
Since causality requires $|w_{\mp}| < 1$, we arrive at the important
conclusion
\begin{equation} \label{48}
{\rm sgn}\, w_{\mp}' = \mp {\rm sgn} \left[ \Sigma~\epsilon' \right]~.
\end{equation}
For a simple rarefaction
wave moving to the right, i.e., where $\epsilon' <0$, the inverse
slope (\ref{ch1}) of adjacent ${\cal C}_-$--characteristics has to increase for
$\Sigma >0$, and to decrease for $\Sigma <0$, and stays constant
for $\Sigma=0$. This means that for TN matter the slope of the
characteristics will decrease, i.e., they
``fan out'' in the $x-t$--plane and the rarefaction proceeds undisturbed as
a simple wave \cite{courant}, i.e., {\em for TN matter the
stable hydrodynamic solution is a simple rarefaction wave}.
In turn, the hydrodynamic flow emerging from the initial condition (\ref{in1},
\ref{in2}) will thus be a single simple rarefaction wave.

On the other hand, for TA matter the slope of the characteristics will
increase, i.e., they are
bound to intersect, which is a signal for the formation of shock waves
\cite{courant}. Thus, {\em for TA matter the stable hydrodynamic solution
is a rarefaction shock wave}.
This also means that the initial discontinuity (\ref{in1}, \ref{in2})
cannot decay.
Since an EoS is usually not completely TA, but contains regions where matter
is also TN (see subsection 4.3), the hydrodynamic flow
will feature discontinuities in TA regions as well
as simple waves in TN regions of the EoS.

For the special case $\Sigma=0$, all characteristics have the
same slope which indicates the existence of a region of constant flow.
If $\Sigma=0$ globally, we would then
simply {\em preserve\/} the initial situation
(\ref{in1}, \ref{in2}) for all times (as discussed above,
the same is true for globally TA matter). Since the EoS we will
finally consider (see subsection 4.3) has TN regions as well,
only {\em that\/} part of the initial discontinuity will be preserved which
corresponds to energy densities where $\Sigma=0$. This means that
even if $\Sigma$ does not
really change sign but only vanishes, {\em for the discontinuous
initial conditions\/} (\ref{in1}, \ref{in2})
{\em we will observe a rarefaction shock wave in the subsequent
hydrodynamical evolution}.

Similar considerations apply for a simple rarefaction wave moving to the
left, where the ${\cal C}_+$--characteristics have to be considered.
The result is again that the hydrodynamically stable solution is a
{\em simple rarefaction wave for TN matter and a rarefaction shock wave
for TA matter}.

For {\em compressional\/} waves, one infers
from (\ref{48}) that the picture reverses and characteristics
are bound to intersect for TN matter and ``fan out'' for TA matter.
Thus, the hydrodynamically stable solution is a {\em compressional shock wave
for TN matter and a simple compression wave for TA matter}. This will be
discussed in detail in \cite{test2}.

\subsection{Thermodynamically normal matter}

Consider the simple EoS
\begin{equation} \label{EoS1}
  p(\epsilon) = c_s^2\, \epsilon~,~~c_s^2  = const.~.
\end{equation}
For $c_s^2 = 1/3$, this is the EoS of an ultrarelativistic ideal gas.
One easily checks that $\Sigma = 2\, c_s^2 (1-c_s^2)/(1+c_s^2)\epsilon >0$,
i.e., this EoS describes TN matter.

As shown in the previous subsection, in TN matter the hydrodynamically
stable solution for the decay of the initial discontinuity (\ref{in1},
\ref{in2}) is a single simple rarefaction wave,
where ${\cal R}_+ =const.$ \cite{courant}.
For the initial conditions (\ref{in1}, \ref{in2}) we determine the value
of ${\cal R}_+$ at $t=x = 0$ to be ${\cal R}_+=0$ and thus
\begin{equation}
y = - \int_{\epsilon_0}^\epsilon \frac{ c_s\,{\rm d}\epsilon' }{\epsilon' +
p(\epsilon')}~.
\end{equation}
For the EoS (\ref{EoS1}) this gives
\begin{equation} \label{vp}
v(\epsilon) = \tanh y(\epsilon) =
\tanh \left[ - \frac{c_s}{1+c_s^2} \ln \left\{ \frac{\epsilon}{\epsilon_0}
\right\} \right] =\frac{1-(\epsilon/\epsilon_0)^{2c_s/(1+c_s^2)}}{
1+(\epsilon/\epsilon_0)^{2c_s/(1+c_s^2)}}~.
\end{equation}
This equation nicely shows the defining
property of a simple wave, namely that there is a unique
relationship between the value of the fluid velocity and its
thermodynamic state \cite{courant}. Furthermore, with the help of
(\ref{ch1}) for ${\cal C}_-$, we can now calculate for any given $\epsilon$,
$0 \leq \epsilon \leq \epsilon_0$, the position $x(t;\epsilon)$
at which this value for the energy density occurs for given $t$,
\begin{equation} \label{x}
x(t;\epsilon) = \frac{v(\epsilon) - c_s}{ 1 - v(\epsilon)~c_s}~t~.
\end{equation}
Eq.\ (\ref{x}) has
{\em similarity\/} form, i.e., the profile of the rarefaction wave
does not change with time when plotted as a function of the
{\em similarity\/} variable $\zeta \equiv x/t$.

Let us now complete the solution of the hydrodynamic problem.
Causality requires that the initial conditions (\ref{in1}, \ref{in2})
remain unchanged for $|\zeta| > 1$. Thus, we only have to determine
the solution in the range $-1 \leq \zeta \leq 1$ (the forward light cone).
 From eqs.\ (\ref{vp}, \ref{x}) we infer that
the head of the simple rarefaction wave (the point where the rarefaction
of matter starts, i.e., where the energy density $\epsilon$
starts to fall below $\epsilon_0$) travels with the velocity $-c_s$
to the left. On the other hand,
the base of the rarefaction wave (the point where the vacuum ends, i.e.,
where $\epsilon$ starts to acquire non-vanishing values) travels with light
velocity $v = 1$ to the right. Thus, the energy density as a function
of $\zeta$ can be written as
\begin{equation}
\epsilon(\zeta) = c_s^{-2} p_0 \cdot \left\{ \begin{array}{ll}
                   1 &,~~-1 \leq \zeta \leq -c_s \\
         {\displaystyle \left[ \frac{1-c_s}{1+c_s}~ \frac{1-\zeta}{1+\zeta}
                   \right]}^{(1+c_s^2)/2c_s} &,~~ -c_s < \zeta \leq 1~.
                             \end{array}  \right.
\end{equation}
The velocity can then be inferred from (\ref{vp}), or simply
from (\ref{x}), and the
other hydrodynamic variables from the EoS (\ref{EoS1}) and (\ref{E}--\ref{R}).
With the thermodynamic relation ${\rm d}p = s~{\rm d}T = (\epsilon +p)~
{\rm d} T/T$ and the EoS (\ref{EoS1})
we get the temperature $T/T_0 = (\epsilon/\epsilon_0)^{c_s^2/(1+c_s^2)}$.
In Fig.\ 1 we show
(a) the time evolution of $T/T_0$, (b) the $T/T_0-$profile
as a function of $\zeta$ (which is time-invariant as explained above),
and (c) the $T^{00}-$profile
normalized to the initial pressure $p_0$ for $c_s^2=1/3$.

\subsection{Thermodynamically anomalous matter}

Consider the EoS for a (net) baryon-free
QGP of gluons, $u$ and $d$ quarks, described
by the MIT bag model \cite{MIT}, and that of an ideal gas of massless pions.
Match both EoS's via Gibbs equilibrium conditions, i.e., equate
pressure and temperature to obtain a first order phase transition
at temperature $T_c$ with pressure $p_c=p_Q(T_c) =p_H(T_c)$. This yields
\begin{equation}
p(T) = p_c \cdot \left\{ \begin{array}{ll}
    r {\displaystyle \left(\frac{T}{T_c}\right)}^4 - B/p_c & ,~~T\geq T_c \\
    {\displaystyle \left(\frac{T}{T_c} \right)}^4  &  ,~~T < T_c~.
               \end{array} \right.
\end{equation}
Here $r=g_Q/g_H$ is the ratio of degrees of freedom in the QGP and
the pion gas ($g_Q=37$ and $g_H=3$ in our case), $B$ is the MIT bag
constant, and $T_c = [90B/\pi^2 g_H (r-1)]^{1/4}$. Note that $B/p_c \equiv
r-1$.
As a function of $\epsilon$,
\begin{equation}
p(\epsilon) = \left\{ \begin{array}{ll}
             ( \epsilon - 4B)/3 & ,~~\epsilon \geq \epsilon_Q \\
             p_c & ,~~\epsilon_Q > \epsilon > \epsilon_H   \\
             \epsilon/3 & ,~~\epsilon_H \geq \epsilon~,
                      \end{array} \right. \label{EoS2}
\end{equation}
where $\epsilon_Q \equiv 3p_c + 4B=(4r-1)p_c$,
$\epsilon_H \equiv 3p_c$. Obviously,
\begin{equation}
c_s^2 = \frac{{\rm d}p}{{\rm d} \epsilon} = \left\{ \begin{array}{ll}
           1/3 & ,~~\epsilon \geq \epsilon_Q \\
           0 & ,~~\epsilon_Q > \epsilon > \epsilon_H   \\
           1/3 & ,~~\epsilon_H \geq \epsilon~,
                      \end{array} \right.
\end{equation}
i.e., from eq.\ (\ref{Sigma}) we infer
\begin{equation}
\Sigma = \left\{ \begin{array}{ll}
           1/3(\epsilon-B) & ,~~\epsilon \geq \epsilon_Q \\
           0 & ,~~\epsilon_Q > \epsilon > \epsilon_H   \\
           1/3\, \epsilon & ,~~\epsilon_H \geq \epsilon~.
                      \end{array} \right.
\end{equation}
Thus, the EoS (\ref{EoS2}) has TN regions above $\epsilon_Q$ and below
$\epsilon_H$. In the mixed phase $\Sigma$ vanishes.
As discussed in subsection 4.1, for the initial conditions
(\ref{in1}, \ref{in2})
this will lead to the formation of a rarefaction shock wave, as occurs in
the expansion of TA matter.

If the initial state (\ref{in1}) has an energy density
$\epsilon_0 \leq \epsilon_H$, matter is TN and the expansion is
exactly the same as discussed in the previous subsection.
If, on the other hand, $\epsilon_0 > \epsilon_Q$, the hydrodynamic solution
is a simple rarefaction wave in the TN QGP part of the EoS,
but as soon as the rarefaction
reaches the phase transition point $\epsilon=\epsilon_Q$ on the wave,
one enters the mixed phase region of the EoS, and the initial discontinuity
will be preserved as a rarefaction shock wave \cite{friman,blaizot,ruuskanen}.
The energy density of matter flowing into this shock is $\epsilon_Q$.
Matter emerges from this shock in a hadronic state which will be
determined below and the rarefaction continues as a simple
wave due to that matter being TN.
Finally, if $\epsilon_Q \geq \epsilon_0 > \epsilon_H$ in the initial state,
rarefaction proceeds directly through the rarefaction shock
from which hadronic matter emerges in a simple rarefaction wave.

Let us now determine the state of hadronic matter emerging from
the rarefaction shock.
In the rest frame of a (space-like)
shock wave, conservation of energy and momentum
can be written in the form \cite{LL6,taub}
\begin{equation} \label{shock}
T^{01} = T^{01}_i ~,~~T^{11} = T^{11}_i~,
\end{equation}
where the subscript $i$ labels quantities in front of the shock.
We denote the enthalpy densities
behind and in front of the shock by
$w \equiv \epsilon + p, \, w_i = \epsilon_i + p_i$
and the corresponding space-like components of the hydrodynamic 4--velocity
by $u=\gamma v,\, u_i=\gamma_i v_i$. Then, defining a
generalized volume for (net) baryon-free
matter, $X \equiv w u^2/w_i u_i^2$ \cite{ruuskanen},
one can derive a Taub equation \cite{taub}
from eqs.\ (\ref{shock}) \cite{ruuskanen}:
\begin{equation} \label{teq}
wX - w_i - (p-p_i) (1 +X) = 0~.
\end{equation}
Another consequence of the conservation laws (\ref{shock}) is the relation
\begin{equation} \label{ratio}
\frac{v_i}{v} = \frac{p_i + \epsilon}{p + \epsilon_i}
\end{equation}
for the velocities of matter going into and coming out of
the shock wave.

Suppose the initial state had $\epsilon_0 \geq \epsilon_Q$. Then, the
rarefaction shock sets in at
$p_i \equiv p_c,\, \epsilon_i \equiv \epsilon_Q$ and $w_i \equiv 4 (p_c + B)
= 4\, r p_c $.
 From the Taub equation (\ref{teq}) one derives \cite{ruuskanen}
that the Taub adiabat $p(X)$ is given by
\begin{equation} \label{px1}
p(X) = p_c \cdot \left\{ \begin{array}{ll}
                    1 & ,~~ 1 \leq X < r \\
        {\displaystyle \frac{ 4r-1-X}{3X-1} }& ,~~ r \leq X \leq 4r-1~,
                   \end{array} \right.
\end{equation}
where the upper line corresponds to final states in the mixed phase and
the lower line to those in the hadronic phase.
Similarly, one can derive a Taub adiabat for the initial state being
in the mixed phase \cite{ruuskanen}:
\begin{equation} \label{px2}
p(X) = p_c \cdot \left\{ \begin{array}{ll}
      1 & ,~~ 1 \leq X < (3E_0+1)/4 \\
     {\displaystyle \frac{3E_0 -X}{3X-1}} & ,~~(3E_0+1)/4 \leq X \leq 3E_0~,
                    \end{array} \right.
\end{equation}
where $E_0 \equiv \epsilon_0 / \epsilon_H$. Again, the upper line
is for mixed phase final states, the lower for hadronic final states.
Fig.\ 2 shows (a) the Taub adiabat (\ref{px1}) and (b) the Taub adiabat
(\ref{px2}) for $E_0 = 5,10$.
Also shown is the Chapman--Jouguet (CJ) point on the respective adiabats
\cite{LL6}. It is defined as the point where the chord from the initial
state $(1,p_c)$ to the final state $(X,p(X))$ has the same slope as the
Taub adiabat.
This point represents final states where the hadronic fluid emerges
with the velocity of sound from the shock and where the entropy
production through the shock is maximized \cite{LL6}.

In principle, the conservation laws (\ref{shock}) allow any state on
the Taub adiabat as final state for the rarefaction shock.
This is in contrast to compression shocks in the initial stage of a heavy--ion
collision where the final compressed state is uniquely determined by the
beam energy of the colliding nuclei (see e.g.\ \cite{test2}).
However, mechanical stability of the shock \cite{LL6,gorenstein} requires that
the chord from the initial state $(1,p_c)$ to the final state $(X,p(X))$
does not intersect the Taub adiabat.
Therefore, final states below the CJ point are not allowed.

Furthermore, not all of the remaining final states above the CJ point
correspond to the long-term hydrodynamic solution, or
in other words, lead to a stationary hydrodynamic profile.
If the final state lies above the CJ point, matter emerges from the shock
at a higher density and, consequently, with a smaller
velocity than at the CJ point, where it is equal to the velocity of sound.
However, the subsequent rarefaction
proceeds via a simple wave where the velocity of matter relative to
the wave profile is the sound velocity, see eq.\ (\ref{ch1}). This means that
e.g.\ for the head of the rarefaction wave (where it is attached to the
shock) the flow velocity will be larger than that of
matter emerging from the shock.
Eventually, this excess of matter flow leads to a density
decrease at the head of the
rarefaction wave, i.e., in the final state of the shock, and
consequently to an increase of the shock strength and the velocity of matter
emerging from the shock. This proceeds until the latter reaches
the velocity of sound and matches the velocity of matter
on the rarefaction wave. Now the solution becomes stationary, i.e.,
the profile of the rarefaction wave no longer shifts with respect
to the shock.
Thus, in a stationary situation, the final state of the shock will
always correspond to the CJ point on the Taub adiabat. Remarkably, this process
satisfies the principle of entropy maximization.

Let us now complete the solution for the expansion of TA matter into vacuum.
We consider first the case when the initial energy density of
semi-infinite matter $\epsilon_0 > \epsilon_Q$. According
to Fig.\ 3 (a), we can distinguish five regions of flow.
Region I extends over $-\infty < x < x_A=-c_s t$ and corresponds to constant
flow with $\epsilon=\epsilon_0,\, v=v_0=0$. Region II starts at $x_A$ and
is a simple rarefaction wave in the QGP which terminates at $x_B=t
(v(\epsilon_Q) -c_s)/(1-v(\epsilon_Q)c_s)$, where
(cf.\ derivation of eq.\ (\ref{vp}))
\begin{equation} \label{veQ}
v(\epsilon) =
\frac{1-[(\epsilon-B)/(\epsilon_0-B)]^{2c_s/(1+c_s^2)}}{
1+[(\epsilon-B)/(\epsilon_0-B)]^{2c_s/(1+c_s^2)}}~.
\end{equation}
Region III is a region of constant flow, $\epsilon=\epsilon_Q, \,
v=v(\epsilon_Q)$ and extends from $x_B$ to $x_C = v_{sh}t$, $v_{sh}$ being the
velocity of the rarefaction shock wave, which will be determined subsequently.
Region IV extends from $x_C$ to $x_D= t$ and represents a simple rarefaction
wave in hadronic matter. At the head of this wave, $\epsilon = \epsilon_{CJ}
\equiv 3\, p_{CJ},\, v=v_{CJ}=(v_{sh}+c_s)/(1+v_{sh}c_s)$ (the pressure
at the CJ point $p_{CJ}$ is given below).
On this wave, application of (\ref{Rpm}) for ${\cal R}_+=0$ yields now
\begin{equation} \label{vp2}
v(\epsilon) =
\tanh y(\epsilon) = \tanh \left[ - \frac{c_s}{1+c_s^2} \ln \left\{
\frac{\epsilon}{\epsilon_{CJ}} \right\} + y_{CJ} \right] ~,
\end{equation}
where $y_{CJ}={\rm Artanh}~v_{CJ}$ is the fluid rapidity of
matter emerging from the shock.
Finally, region V corresponds to vacuum, $\epsilon=0$, $v=1$.

To determine $v_{sh}$ we use eq.\ (\ref{ratio}),
\begin{equation} \label{vi}
\frac{v_i}{c_s} = \frac{p_c + \epsilon_{CJ}}{p_{CJ} +
\epsilon_Q} = \frac{1+3p_{CJ}/p_c}{4r-1 + p_{CJ}/p_c}~,
\end{equation}
where the pressure at the CJ point is given by
\begin{equation} \label{pCJ}
p_{CJ} = p_c~\frac{3r-1 - \sqrt{r^2 -(4r-1)/3} }{3r-1 + 3 \sqrt{r^2
-(4r-1)/3}}~.
\end{equation}
In the rest frame of undisturbed matter, the velocity of matter flowing
into the shock is $v(\epsilon_Q)$, thus, the shock velocity in this frame is
given by
\begin{equation}\label{vsh}
v_{sh} = \frac{v(\epsilon_Q)-v_i}{1-v(\epsilon_Q)v_i}~.
\end{equation}
We observe that the solution is again of similarity type. Thus,
the energy density as a function of the similarity variable $\zeta$
reads (in the forward light cone)
\begin{equation}
\epsilon(\zeta) = p_c \cdot \left\{ \begin{array}{ll}
          \epsilon_0/p_c & , ~~-1 \leq \zeta \leq -c_s \\
{\displaystyle \left[\frac{1-c_s}{1+c_s}~\frac{1-\zeta}{1+\zeta}\right]^{(1+
c_s^2)/2c_s}\left(\frac{\epsilon_0}{p_c} -r + 1\right)
 + r-1}  &,~~ -c_s < \zeta \leq {\displaystyle \frac{v(\epsilon_Q)-c_s}{
1-v(\epsilon_Q)c_s}} \\
         4r-1& ,~~{\displaystyle \frac{v(\epsilon_Q)-c_s}{1-v(\epsilon_Q)c_s}}
< \zeta \leq  v_{sh} \\
{\displaystyle \left[\frac{1-c_s}{1+c_s}~\frac{1-\zeta}{1+\zeta}~
\frac{1+v_{CJ}}{1-v_{CJ}} \right]^{(1+c_s^2)/2c_s}\frac{\epsilon_{CJ}}{p_c}} &
 ,~~ v_{sh}< \zeta \leq 1~.
                             \end{array}  \right.
\end{equation}
Other hydrodynamic variables can be obtained from the EoS (\ref{EoS2}),
eqs.\ (\ref{veQ}, \ref{vp2}), and (\ref{E}--\ref{R}).
The temperature follows from
\begin{equation} \label{ttc}
\frac{T}{T_c} = \left\{ \begin{array}{ll}
 {\displaystyle \left(\frac{\epsilon-B}{\epsilon_Q-B}\right)^{c_s^2/
(1+c_s^2)} } &,~~ \epsilon \geq  \epsilon_Q \\
 ~~~1 &,~~ \epsilon_Q > \epsilon > \epsilon_H \\
 {\displaystyle \left(  \frac{\epsilon}{\epsilon_H}\right)^{c_s^2/
(1+c_s^2)} } &,~~ \epsilon_H \geq  \epsilon~.
                        \end{array} \right.
\end{equation}
In Fig.\ 4 we show
(a) the time evolution of $T/T_c$, (b) the $T/T_c-$profile
as a function of $\zeta$, and (c) the $T^{00}-$profile normalized to
the critical pressure $p_c$ for an initial energy density
$\epsilon_0 = 603 \frac{1}{3}\, p_c$. This corresponds to an
initial temperature $T_0/T_c= 2$. Figs.\ 4 (d--f) shows the same for
$\epsilon_0 = 50\, p_c$ which
is only slightly above $\epsilon_Q= (4r-1)p_c=48 \frac{1}{3}\, p_c$,
with an initial temperature $T_0/T_c \simeq 1.011$.

Let us now construct the solution for the initial state being
in the mixed phase. According to Fig.\ 3 (b), we can distinguish
three different regions.
Region I is constant flow with $\epsilon = \epsilon_0=E_0 \epsilon_H$,
$v=0$ and extends over $-\infty < x \leq x_A=v_{sh}t$. Since
the velocity of undisturbed matter is zero, the (modulus of the) velocity
$v_{sh}$ of the rarefaction shock travelling into that region is identical to
the velocity $v_i$ of matter streaming into this shock in its rest frame.
With the help of (\ref{ratio}) this leads to
\begin{equation} \label{vsh2}
v_{sh} \equiv - v_i = - c_s~\frac{1+3 p_{CJ}/p_c}{3E_0 + p_{CJ}/p_c}~,
\end{equation}
where now
\begin{equation} \label{pCJ2}
p_{CJ} = p_c~\frac{9E_0 -1 - \sqrt{9E_0^2 - 10E_0 +1}}{9E_0-1+3
\sqrt{9E_0^2 - 10E_0 +1}}~,
\end{equation}
and as before $E_0 = \epsilon_0/\epsilon_H \equiv \epsilon_0/3p_c$.
Region II is a simple rarefaction wave in hadronic matter
and extends from $x_A$ to $x_B=t$.
Finally, region III corresponds to vacuum, $\epsilon=0,\, v=1$.
In terms of $\zeta$, the solution reads
\begin{equation}
\epsilon(\zeta) = p_c \cdot \left\{ \begin{array}{ll}
          3 E_0 & , ~~-1 \leq \zeta \leq v_{sh} \\
{\displaystyle \left[\frac{1-c_s}{1+c_s}~\frac{1-\zeta}{1+\zeta}~
\frac{1+v_{CJ}}{1-v_{CJ}} \right]^{(1+c_s^2)/2c_s}\frac{\epsilon_{CJ}}{p_c}} &
 ,~~ v_{sh}< \zeta \leq 1~.
                             \end{array}  \right.
\end{equation}
Here, $\epsilon_{CJ} = 3\, p_{CJ}$ with $p_{CJ}$ now given by (\ref{pCJ2}),
$v_{sh}$ is given by (\ref{vsh2}), and again $v_{CJ} =
(v_{sh}+c_s)/(1+v_{sh}c_s)$. We will not consider numerical
solutions for $\epsilon_0$ in the mixed phase, and thus do not show
analytical results as in Fig.\ 4 for the previous case.
The reason is that the situation is very similar
to that in Figs.\ 4 (d--f) (see also discussion in Section 6)
and does not exhibit new insights from the analytical or
numerical point of view.

In a sense, the two cases shown in Fig.\ 4
represent opposite scenarios, both from the physical as well as from the
numerical point of view. While the first might be the appropriate initial
condition for a QGP created at RHIC or LHC, the second is more likely to
resemble initial conditions at the CERN--SPS. The first
features a very fast hadronizing rarefaction shock, the second a very slow,
stationary one. This is illustrated in Fig.\ 5 where we
show the shock velocity as given by (\ref{vsh}, \ref{vsh2})
as a function of the initial energy density $\epsilon_0$.
As one observes, for a range of initial energy densities around the boundary
of mixed phase and QGP the shock has small velocities in the
rest frame of undisturbed matter.
This has interesting consequences
for the lifetime of the mixed phase in heavy--ion collisions \cite{shuryak},
see also Section 6.

\section{The one--dimensional expansion into vacuum -- numerical solution}

In this section we compile numerical results for
the expansion of semi-infinite
matter into vacuum. In Fig.\ 6 we show the
expansion of TN matter described by the EoS (\ref{EoS1}) for the relativistic
HLLE with $\lambda = 0.99$ and for the SHASTA with
$\lambda = 0.4$. The grid spacing is taken to be $\Delta x=1$ for both cases.
Figs.\ 6 (a,d) show the numerically calculated $T/T_0$--profile
for 0, 10, 20,..., 100 time steps, while the corresponding parts (b,c,e,f)
show $T/T_0$ and $T^{00}/p_0$ as a function of $\zeta$
after 5, 10, 20 and 50 time steps
(for the SHASTA also after 100 time steps) as compared to the analytical
result of Figs.\ 1 (b,c).
The analytical solution is remarkably well reproduced, up to
effects due to numerical viscosity.
The SHASTA shows a tendency to produce small fluctuations around the
point where the profile is stationary ($v \simeq c_s$). As a long-term
run has proved, these fluctuations are not instabilities. They can
be removed by reducing the antidiffusion, see Appendix for details.

The advantage of plotting quantities as functions of the similarity
variable $\zeta$ becomes obvious from Figs.\ 6 (b,c,e,f):
curves for different times should all lie on top of each other. Thus,
one can monitor the approach to the analytical solution in an
elegant and compact way.
One observes that this approach is rather fast for the HLLE
(about 20 time steps)
and slightly slower for the SHASTA (about 50 time
steps). Of course, since the rarefaction wave spreads
over a growing number of cells (which is linear
proportional to the elapsed time),
the trivial effect of a {\em better resolution\/} on the $\zeta$--axis
is partly responsible for this approach.
In this sense, for the same $\Delta x$
the relativistic HLLE with the larger CFL number $\lambda$
has an advantage over the SHASTA:
the physical time has advanced further after the
same number of time steps than in the SHASTA and, consequently, the
similarity wave profile extends over a larger number of cells. This can
also be seen comparing the mutual distance of symbols
(time steps 5, 10, and 20) in
Fig.\ 6 (b,c) with that in (e,f). Taking this effect into account, we
conclude that the approach of the relativistic HLLE and the
SHASTA to the analytical solution takes approximately {\em the same
physical time\/} $t \simeq
20\, \lambda_{HLLE} \Delta x \simeq 50\, \lambda_{SHASTA} \Delta x
\sim 20\, \Delta x$.

Fig.\ 7 shows the expansion of semi-infinite TA matter for an initial
temperature $T_0/T_c = 2$ (cf.\ Figs.\ 4 (a--c)) for the relativistic
HLLE and the SHASTA,
respectively. The reproduction of the complicated hydrodynamic
solution is quite good for the HLLE and satisfactory for the SHASTA,
although the latter has, after 100 time steps, still
problems separating the simple rarefaction wave in
the hadronic phase from the rarefaction shock and again
shows wiggles where the profile is stationary. However, as
in Fig.\ 6, the approach to the
analytical solution takes about the same {\em physical\/} time for {\em both\/}
algorithms.

In Fig.\ 8 the performance of relativistic HLLE and SHASTA
is investigated for the initial condition of Figs.\ 4 (d--f).
While the approximation of the analytical solution is
quite good after 100 time steps for both algorithms, the
SHASTA performs globally better for
this problem than the HLLE. In particular, it approaches the analytical
solution after a {\em shorter\/} physical time.

In Fig.\ 8 (b) it seems that the HLLE algorithm places
the rarefaction shock wave at too large values of $\zeta$ in the
first few time steps and approaches the correct position only for later
times. Inspecting Fig.\ 8 (c), however, reveals that this interpretation
is misleading. Here the discontinuity is correctly placed, but
smeared out due to numerical diffusion.
While this smearing is symmetric around the true position of
the discontinuity in the energy density, it appears asymmetric in the
temperature, because this quantity is proportional to the fourth
root of the energy density, cf.\ eq.\ (\ref{ttc}).

To quantify the deviation from the analytical solution we plot in Fig.\ 9
the quantity
\begin{equation} \label{dev}
d \equiv \int_{-2}^{3} {\rm d}\zeta \left[ T^{00}_{num}(\zeta) -
T^{00}_{ana} (\zeta) \right]^2
\end{equation}
as a function of the number of time steps. (The integral is
approximated by a Riemann sum, $\int {\rm d}x~f(x) \simeq
\sum_i f(x_i)(x_{i+1}-x_i)$). We also measure the deviation separately
in the acausal region $-2 < \zeta <-1,\, 1 < \zeta \leq 3 $,
in order to determine the prediffusion.
The corresponding part of the deviation (\ref{dev}) is the dotted
line in Fig.\ 9.

The {\em absolute\/}\footnote{There is no reasonable way to define
a {\em relative\/} deviation due to the occurrence of vacuum.}
measure of deviation (\ref{dev}) confirms the
observations made in Figs.\ 6--8, namely that
the HLLE is more accurate than the SHASTA, except for
the expansion problem of Fig.\ 8.
For the problems of Figs.\ 6, 7 this holds even when comparing the deviation at
the same {\em physical\/} time.

Furthermore, for the HLLE the prediffusion is essentially zero up to
50 time steps.
At this time, the prediffusion has just ``leaked'' half a cell size
($50 (1- \lambda) \Delta x \equiv \Delta x/2$)
into the acausal region. The astonishing fact is that, despite the rather
disadvantageous $\lambda=0.4$, the prediffusion in the SHASTA is
remarkably small (cf.\ also the profiles in Figs.\ 6--8). A calculation
with the HLLE for the same $\lambda = 0.4$ gives a much larger deviation due to
prediffusion than that for the SHASTA.
Note that the complexity of the expansion solution for TA matter as compared
to TN matter has caused to increase $d$ by several orders of magnitude
in Figs.\ 9 (b,c) as compared to Fig.\ 9 (a). However, since $d$ is not a
relative measure
for the deviation from the analytical solution, it can only serve
for mutually comparing the performance of different algorithms.

The above investigation has shown that the algorithms are well
able to handle the vacuum and rarefaction shock waves in the
expansion of TA matter. However, they need a certain
number of time steps before they approach the analytical solution.
If one wants to limit the initial deviation to a given physical
time interval one has to choose a
sufficiently small time step width $\Delta t$, or, for fixed $\lambda$,
a sufficiently fine spatial resolution $\Delta x$.
For instance, if the numerical reproduction of the hydrodynamical solution is
taken to be sufficiently accurate after 50 time steps and one wants this
accuracy to be achieved after a physical time of 1 fm,
one is bound to choose $\Delta t=1\,{\rm fm}/50=0.02$ fm, or
$\Delta x = 0.02\,{\rm fm}/ \lambda$.
A system with longitudinal extension $L$ must then be
resolved over $50\,\lambda L/$fm grid points.
This requires rather large grids in order to study
dynamical problems over longer periods of time.

Let us finally mention that in the HLLE runs
we do not use the physical velocity of sound in the
mixed phase ($c_s \equiv 0$) in the signal velocity estimates (\ref{signal1},
\ref{signal2}). We found
that doing so prevents matter transport across the rarefaction
shock if the velocity of the latter is small and leads
to creation of vacuum between rarefaction shock and hadronic simple wave.
Rather, we {\em globally\/} take $c_s^2 = 1/3$ in the signal velocity estimates
which is safe but also increases the numerical dissipation.
Furthermore, in the actual SHASTA propagation, we take $v=0$ for vacuum
cells, as is also necessary for multidimensional applications,
cf.\ footnote 6. We checked that the difference to the choice $v=1$
as in (\ref{in2}) is negligible anyway.

\section{Expansion in finite systems}

In this section we discuss the one--dimensional expansion
into the vacuum for finite systems, i.e., we start with the initial
condition
\begin{eqnarray} \label{in3}
\epsilon (x,0) & = & \left\{ \begin{array}{ll}
                 \epsilon_0 &,~~|x| \leq R \\
                          0 &,~~|x| > R~,
                            \end{array} \right. \\
v(x,0) & = & \left\{ \begin{array}{ll}
                         -1 &,~~-\infty < x < - R \\
                          0 &,~~-R \leq x \leq R \\
                          1 &,~~R <x < \infty~,
                            \end{array} \right. \label{in4}
\end{eqnarray}
where $2R$ is the size of the system ($R$ is the ``radius'').
Let us briefly sketch the subsequent hydrodynamical evolution.
In Section 4 we have seen that a simple rarefaction wave travels into
undisturbed TN matter with velocity $v_{rare} = c_s$.
For an initial state in the mixed phase,
the rarefaction is accomplished by a shock with velocity
$v_{rare}=v_{sh}$ as given by (\ref{vsh2}). Thus, for times
$t < R/ v_{rare}$, i.e., as long as the rarefaction has not reached
the center of the system at $x=0$, the evolution of the system for $x >0$
is exactly the same as described in Section 4, and for $x<0$
simply the reflected version of that. In this stage, the solution
is still of similarity type. However, as soon as the two rarefaction
waves reach the center and start to overlap, the solution becomes more
complicated and is no longer of similarity type.

Nevertheless, as was shown by Landau \cite{landau}, if matter
has an EoS of the form (\ref{EoS1}), an analytical solution still exists.
Its derivation is lengthy and shall not be repeated here.
The general structure, however, is the following:
in the {\em center\/} of the system where
the (in this case, simple) rarefaction waves overlap,
the hydrodynamic solution is determined by a potential $\chi(T,y)$.
As matter expands, the overlap region grows, but since it can
do so only causally, {\em outside\/} this region
the hydrodynamic solution remains a simple rarefaction
wave. The boundary between the two regions is determined by
smoothly matching the hydrodynamic flow.

A practical way to generate a semi--analytic solution to this problem
{\em in the whole forward light cone\/} is via the method of
characteristics \cite{baym}.
The basic idea is to solve the system of four characteristic equations
(\ref{ch2}, \ref{ch1}) which are ordinary differential equations.
This method was described in detail in \cite{baym} and shall not be repeated
here. We will use it to compare with the results produced by the algorithms of
Section 3. This method, however, is also
restricted to TN matter, since the characteristics will intersect at
shock discontinuities \cite{courant}, as occur in the expansion of TA matter.
Then, the initial value problem posed by (\ref{ch2}, \ref{ch1}) is no longer
unique and thus not solvable.

In Figs.\ 10, 11 we show the results for the expansion of TN matter
with $c_s^2=1/3$ for the
relativistic HLLE and the SHASTA in comparison to the semi-analytic
result generated by the method of characteristics \cite{baym}. We again choose
$\lambda=0.99$ for the HLLE and $\lambda = 0.4$ for the SHASTA.
Of special interest is the dependence of the results on the
spatial resolution:
Figs.\ 10, 11 (a,b) show the temperature and (c,d) the
energy density for $\Delta x = 0.1\, R$ and
$\Delta x = 0.025\, R$, respectively. One observes that in the first case
both quantities are overestimated in the central region
at intermediate times and also that viscosity effects are more prominent.

This is explained as follows: as the results of the previous
section have indicated, it takes about 20 time
steps for the HLLE and of the order of 50 for the SHASTA
to correctly approximate the analytical shape of a simple
rarefaction wave. This criterion is not fulfilled in Figs.\ 10, 11 (a,c),
since the simple rarefaction wave reaches the center after about 14 time steps
in Figs.\ 10 (a,c) and about 30 time steps in Figs.\ 11 (a,c). The
two rarefaction waves then start to overlap before they were able to
assume their correct shape. Subsequently, the numerics fail
to reproduce the correct behaviour in the overlap region,
leading to the overestimate of temperature and energy density in that
region.

On the other hand, in Figs.\ 10 (b,d) the simple rarefaction wave
travelled for about 60 time steps (HLLE) and in Figs.\ 11 (b,d) for
about 130 time steps (SHASTA) before overlapping with its partner. It
thus had sufficient time to establish the correct shape.
Consequently, the analytical values for quantities in the overlap
region are much better approximated by the numerics.
As in the expansion of semi-infinite matter,
the relativistic HLLE performs better than the SHASTA.
We note that in all cases energy and momentum is globally conserved on the
level of the calculational accuracy (i.e., $10^{-15})$.

 From the physical point of view we would like to draw attention to the
phenomenon that for fixed time $t$ in the global rest frame of the system,
matter at small $|x|$ in the overlap region of the two rarefaction
waves is colder and less dense than for larger $|x|$.
This is a purely relativistic
effect: since the matter velocity is a monotonously increasing function
of $|x|$, and since time in the local rest frame of fast matter
is dilated as compared to that of slow matter, at a fixed global $t$
the rarefaction for cells at small $|x|$ has already proceeded further
than for cells at larger $|x|$. This leads to a
localized wave of denser material travelling outward
which is best seen in the energy density profiles, Figs.\ 10, 11 (c,d).

In Fig.\ 12 we show the expansion of matter described by the
EoS (\ref{EoS2}) for the relativistic HLLE and in Fig.\ 13 for
the SHASTA.
Part (a) and (c) show temperature and energy density, respectively,
for $\epsilon_0=603 \frac{1}{3}\, p_c$,
while part (b) and (d) show these quantities for
$\epsilon_0 = 50\, p_c$. We choose $\Delta x =0.025\, R$,
which, according to the results of Section 5, allows for a reasonable
reproduction of the analytic similarity solution before overlap occurs.
In order to better distinguish the profiles for different times we
show them alternatingly as full and dotted lines.
As long as the rarefaction waves do not overlap, the codes reproduce
the features of the expansion of semi-infinite TA matter discussed
in Sections 4, 5. The SHASTA again tends to produce small ripples
at the stationary point of the QGP rarefaction profile for the case
$T_0/T_c=2$,
see Figs.\ 13 (a,c), which can again be suppressed by
reducing the antidiffusion. This also removes the instabilities on the
hadronic rarefaction wave visible in Fig.\ 13 (a).
Global energy--momentum conservation is again fulfilled within
calculational accuracy.

Figs.\ 12, 13 give a clear picture of the difference in the physical
scenario if we initialize a very hot system ($T_0/T_c \sim 2$) or a system near
the phase transition temperature ($T_0/T_c \sim 1$).
Suppose $\epsilon_0 > \epsilon_Q$. Then,
the simple rarefaction wave accelerates QGP matter until the rarefaction
shock sets in when $\epsilon = \epsilon_Q$. The final hadronic
state is always given by the CJ point (\ref{pCJ}). The important point is
that initial and final state of the shock are {\em independent\/} of
$\epsilon_0$. Thus, the velocity of the
shock {\em in the rest frame of matter moving into the shock\/}
is {\em independent\/} of $\epsilon_0$.

However, for large $\epsilon_0$, the acceleration of matter {\em in front\/} of
the shock is large, so that the latter is explosively driven
{\em outwards\/} by the expanding matter.
On the other hand, for $\epsilon_0$ not much larger than $\epsilon_Q$,
the acceleration of QGP in
front of the shock is small, and consequently, the velocity of the
latter is small or even negative (cf.\ Fig.\ 5). The system
``burns out'' from the inside and seems to ``evaporate'' hadrons.
In both cases, the energy density in the interior decreases until $\epsilon$
drops below $\epsilon_H$ and hadronization proceeds also from the inside.
As explained above, for a fixed time in the global rest frame
of the system, this happens in general earlier for cells at small $|x|$.

Let us mention that for $\epsilon_0$ in the mixed phase, there is no
pre-acceleration by
a simple rarefaction wave and the rarefaction shock ``burns up''
mixed matter with a velocity $v_{sh}$ (given by eq.\ (\ref{vsh2}))
that is the larger the smaller $\epsilon_0$
is. For $\epsilon_0=\epsilon_H$, this velocity reaches $-c_s$ (cf.\
Fig.\ 5) and the rarefaction shock degenerates to a
simple rarefaction wave.
There is no hadronization from the inside in this case since mixed phase matter
is at rest and simply vanishes as soon as the two rarefaction shocks
coming in from right and left meet at the center.

This poses the question whether a remnant of the QGP
survives longer if the system is initially very hot or rather, if it
starts its evolution near the phase transition.
In more physical terms one could ask what is the optimum beam
energy for heavy--ion collisions to prepare a quark--gluon system
with a maximum lifetime \cite{shuryak}.
In \cite{shuryak} it was shown in a three--dimensional calculation (employing
cylindrical symmetry in two space dimensions)
that the mixed phase survives longer if the system is initialized
in this phase than in the hot QGP phase. This behaviour was explained
by the existence of
a so-called ``softest point'', i.e., a minimum in
$p/\epsilon$ as a function of $\epsilon$, in the mixed phase region of the EoS.
The reduction of the pressure relative to the energy density was held
responsible for the observed small flow velocities of mixed phase matter.
Thus, in the rest frame of the system, mixed phase matter does not
expand and is only slowly consumed by ``evaporating'' hadronic matter
with high velocity from its surface (see Fig.\ 2 of Ref.\ \cite{shuryak}).

 From our above investigations we can sustain this picture with a
more detailed understanding of
the slow burning of mixed phase matter and the ``evaporation'' of
fast hadronic matter as being due to a rarefaction shock wave.
To answer the question about the maximum lifetime of the mixed phase we show
in Figs.\ 14 (a,b) isotherms for a temperature $T_c - \delta$ (or equivalently,
curves of constant $\epsilon =\epsilon_H-\delta'$, $\delta,\delta'$
arbitrarily small), as calculated
with the relativistic HLLE (each dot corresponds to a single cell
with the respective temperature).
One observes that for fixed time, the ``burning'' of the mixed phase
proceeds exclusively from the outside for $\epsilon_0$ near $\epsilon_Q$
while for larger $\epsilon_0$ hadronization can also
occur from the inside. This has the consequence that, in agreement with
the conclusions of \cite{shuryak}, at least
at $x=0$ (around midrapidity) the lifetime of the mixed phase is shortened
by the rarefaction following an ``explosive'' expansion scenario and is
prolonged in a ``slow burning'' scenario. (The apparent long lifetime of
the mixed phase at {\em finite\/} $x$ as observed for
$\epsilon_0 \geq 100\, p_c$
is again the mentioned relativistic time dilation effect: for fixed $t$
the {\em proper time\/} of the
corresponding fluid elements has simply not very much advanced due
to their high velocity in the calculational frame; therefore,
they are still hotter.)

However, this is only true up to a certain initial energy density: if
$\epsilon_0$ is very large, it also takes longer to cool the interior and
the lifetime starts to grow again. To be specific we plot in Fig.\ 14 (c)
the time $t_{final}$ where the mixed phase ceases to exist at $x=0$.
For $\epsilon_Q \geq \epsilon_0 \geq \epsilon_H$,
mixed phase matter is consumed by a rarefaction shock with constant
velocity $v_{sh}$ given by (\ref{vsh2}). Thus, the $(T_c-\delta)$--isotherm can
be analytically calculated: it is simply a triangle with its tip located at
$x=0$ and $t\equiv t_{final}/R=1/|v_{sh}|$. Since $|v_{sh}|$ decreases when
$\epsilon_0$ approaches $\epsilon_Q$ (cf.\ Fig.\ 5), the lifetime increases
until $\epsilon_0 = \epsilon_Q$.
For larger $\epsilon_0$ there is no longer an analytical solution and we
have to read off the lifetime from Fig.\ 14 (a) (diamonds in
Fig.\ 14 (c)). However, for {\em very\/}
large $\epsilon_0$ we expect the hydrodynamical behaviour at $x=0$ to resemble
that of Bjorken's solution \cite{bjorken}, i.e., $t_{final} \sim
\epsilon_0^{3/4}$. Such a behaviour is given by the dotted line in Fig.\ 14 (c)
and indeed confirms our expectation.

As one observes, there is a local maximum in the lifetime of the mixed phase,
as predicted in \cite{shuryak}, but asymptotically the lifetime grows
indefinitely proportional to $\epsilon_0^{3/4}$.
In the realistic three--dimensional case, the additional spatial
dimensions will probably only affect the global time scale of cooling, and
thus we expect this picture to remain qualitatively correct.

\section{Conclusions}

In this work we have presented two numerical schemes, the
SHASTA and the relativistic HLLE, to solve the
equations of ideal relativistic hydrodynamics. These algorithms
can be applied to the
simulation of heavy--ion collisions in realistic 3+1--dimensional
space--time geometry, i.e., without assuming specific symmetries to
simplify the solution.
We anticipate that they will find application to analyze
recent data in the BEVALAC energy range \cite{rai}, in order
to understand to what extent ideal relativistic hydrodynamics describes
heavy--ion collisions from the initial compression stage to the final
freeze--out. Furthermore, they are useful as underlying transport
algorithms for multi--fluid-dynamics \cite{katscher} which is
applicable in the AGS to SPS energy range \cite{adrian}.
Finally, they are applied to study the dynamics of
spatial inhomogeneities created in the central region of
ultrarelativistic heavy--ion collisions at RHIC energies \cite{mgdhr}.

We have in detail presented the analytical solution for the one--dimensional
expansion into vacuum, both for
thermodynamically normal as well as thermodynamically anomalous matter.
We have shown that the algorithms
reproduce the analytical solution rather well
after a certain number of time steps.
Choosing a sufficiently small spatial grid size $\Delta x$,
the {\em physical time\/} until the analytical solution is reproduced can
in principle be made arbitrarily small. In realistic situations, this will be
ultimately limited only by the available computational power.

For the one--dimensional expansion of finite systems we have investigated
the lifetime of a mixed phase of pions and QGP.
As expected from scaling hydrodynamics,
for sufficiently large initial energy density $\epsilon_0$
this lifetime grows $\sim \epsilon_0^{3/4}$. However, the lifetime
has also a local maximum near $\epsilon_Q$,
i.e., around the phase boundary between QGP and mixed phase matter which
opens the opportunity to study a long-lived system of deconfined matter
not only in the ultrarelativistic energy domain but probably also
by tuning the beam energy of presently available accelerators \cite{shuryak}.
Let us nevertheless note that in order
to detect this long-lived system via e.g.\ its electromagnetic radiation,
not its lifetime at $x=0$, nor the size of its space--time volume is
most decisive, but rather its initial temperature: the number of emitted
photons or dileptons is only linear proportional to the space--time volume,
but grows proportional to the fourth power of the temperature.
A hot system is more efficiently ``outshining'' other sources of
electromagnetic radiation than a large, but relatively cool system.
\\ ~~ \\
\noindent
{\bf Acknowledgements}
\\ ~~ \\
We thank A.\ Dumitru, M.\ Gyulassy, and U.\ Katscher for stimulating
discussions, and A.\ Dumitru and M.\ Gyulassy
for a careful reading of the manuscript and valuable comments.
Special thanks go to M.\ Gyulassy whose continuous interest and encouragement
contributed essentially to the completion of this work.
\appendix
\section{Appendix}

This appendix serves to illustrate the effect of modifications of
the SHASTA.
We first demonstrate why the phoenical SHASTA
(defined by the antidiffusion fluxes (\ref{Aph})) with the
simplified source treatment (\ref{simsource}) is preferable to
the phoenical SHASTA
with the original sources (\ref{orig})
and the explicit SHASTA (with the antidiffusion fluxes (\ref{Aex}))
with either original or simplified sources.
In Fig.\ 15 we show results for the expansion of (a--c) TN matter and (d--f)
TA matter with $\epsilon_0 = 50\, p_c$ as calculated with these
alternative versions. As one observes in Fig.\ 15 (a),
the phoenical SHASTA produces a ``terrace'' near the stationary
point of the simple wave profile. The explicit SHASTA versions perform
satisfactorily for this expansion problem (Figs.\ 15 (b,c)). However,
all three versions fail to produce the correct profiles for
the expansion of TA matter, see Figs.\ 15 (d--f).

We furthermore show how one can suppress ``wiggles'' on the
stationary point of the rare\-faction wave profile as produced by the
SHASTA version used in the main part of this work, cf.\ Figs.\ 16 (a,b).
To this end, we simply reduce the antidiffusion fluxes (\ref{Aph})
by replacing the first factor 1/8 by a smaller number, e.g. 1/10.
The result is shown in Figs.\ 16 (c,d). The effective reduction of
the antidiffusion by 20\% causes the wiggles to disappear, but, of course,
it also increases the numerical diffusion, especially visible in the larger
prediffusion into the vacuum. Nevertheless, such a reduction
of the antidiffusion fluxes
proves to be crucial to stabilize the transport algorithm in two
space dimensions for large times \cite{mgdhr}.
\\ ~~ \\

\newpage
\noindent
{\bf Figure Captions:}
\\ ~~ \\
{\bf Fig.\ 1:}
Analytic solution for the expansion of TN matter with EoS (\ref{EoS1}).
(a) Temperature normalized to the initial temperature as function of
the spatial variable $x$ for times $t=0,10,20,...,100$. Since the
problem is of the similarity type, $x$ and $t$ carry no units and
can be scaled with an arbitrary, constant factor. (b) $T/T_0$,
and (c) energy density $T^{00}$ normalized
to the initial pressure $p_0$ as functions of the similarity variable $\zeta$.
\\ ~~ \\
{\bf Fig.\ 2:}
(a) The Taub adiabat (\ref{px1}).
(b) The Taub adiabat (\ref{px2}) for $E_0=5$ (solid line) and $E_0= 10$
(dotted line). The corresponding Chapman--Jouguet points and the
chord connecting them with the center $(X,p)=(1,p_c)$ of the adiabat
are also shown.
\\ ~~ \\
{\bf Fig.\ 3:}
Schematic rarefaction profile in TA matter for initial states
(a) $\epsilon_0 > \epsilon_Q$ and (b) $\epsilon_Q \geq \epsilon_0
>\epsilon_H$. See text for a detailed discussion of the different
regions.
\\ ~~ \\
{\bf Fig.\ 4:}
Analytic solution for the expansion of TA matter with EoS (\ref{EoS2})
for (a--c) $\epsilon_0 = 603 \frac{1}{3}\, p_c$ and (d--f)
 $\epsilon_0 = 50\, p_c$.
(a,d) $T/T_c$ as function of $x$ for times $t=0,10,20,...,100$, (b,e) $T/T_c$,
and (c,f) $T^{00}/p_c$ as functions of the similarity variable $\zeta$.
\\ ~~ \\
{\bf Fig.\ 5:}
Velocity of the rarefaction shock in the rest frame of undisturbed matter
as a function of the initial energy density $\epsilon_0$ (in units of
$p_c$).
\\ ~~ \\
{\bf Fig.\ 6:}
Numerical solution for the expansion of TN matter with EoS (\ref{EoS1}),
generated with (a--c) the relativistic HLLE ($\Delta x = 1,\, \lambda = 0.99$)
and (d--f) the SHASTA ($\Delta x = 1,\, \lambda = 0.4$).
(a,d) $T/T_0$ as function of $x$ for 0, 10, 20,..., 100 time steps,
(b,e) $T/T_0$, and (d,f) $T^{00}/p_0$
as functions of $\zeta$ after 5 (stars), 10 (squares), 20 (diamonds),
50 time steps (dotted line) in comparison to the analytical solution (solid
line). The
dashed line in (e,f) corresponds to the profile after 100 time steps.
\\ ~~ \\
{\bf Fig.\ 7:}
Numerical solution for the expansion of TA matter with EoS (\ref{EoS2})
for $\epsilon_0 = 603 \frac{1}{3}\, p_c$,
generated with (a--c) the relativistic HLLE ($\Delta x = 1,\, \lambda = 0.99$)
and (d--f) the SHASTA ($\Delta x = 1,\, \lambda = 0.4$).
(a,d) $T/T_c$ as function of $x$ for 0, 10, 20,..., 100 time steps,
(b,e) $T/T_c$, and (c,f) $T^{00}/p_c$
as functions of $\zeta$ after 5 (stars), 10 (squares), 20 (diamonds),
50 (dotted line), 100 time steps (dashed line) in
comparison to the analytical solution (solid line).
\\ ~~ \\
{\bf Fig.\ 8:}
As in Fig.\ 7 for $\epsilon_0 = 50\, p_c$.
\\ ~~ \\
{\bf Fig.\ 9:}
(a--c) Deviation (\ref{dev}) for the expansion problems of Figs.\ 6--8.
Solid lines are the total deviation, dotted lines the deviation outside
the causal light cone; circles correspond to the relativistic HLLE, squares
to the SHASTA.
\\ ~~ \\
{\bf Fig.\ 10:}
Expansion of a finite system of TN matter with EoS (\ref{EoS1}),
calculated with the relativistic HLLE ($\lambda=0.99$). (a,b)
$T/T_0$-- and (c,d) $T^{00}/p_0$--profiles for times $t=n\lambda R$,
$n=0, 1, 2,..., 10$. For (a,c) $\Delta x = 0.1\, R$, for (b,d) $\Delta x =
0.025\, R$. In (a,c) each dot corresponds to a cell (in (b,d)
a similar presentation is not possible since individual dots would not
be resolved on the scale of the picture; we therefore choose to
draw the profiles as dotted lines).
\\ ~~ \\
{\bf Fig.\ 11:}
As in Fig.\ 10, calculated with the SHASTA ($\lambda = 0.4$).
\\ ~~ \\
{\bf Fig.\ 12:}
Expansion of a finite system of TA matter with EoS (\ref{EoS2}),
calculated with the relativistic HLLE ($\Delta x = 0.025\, R,\,
\lambda = 0.99$). (a,b) $T/T_c$-- and (c,d) $T^{00}/p_c$--profiles for
times $t=n\lambda R$, $n=0, 1, 2,..., 10$ (alternatingly
shown as full and dotted lines). For (a,c) $\epsilon_0=603
\frac{1}{3}\, p_c$, for
(b,d) $\epsilon_0=50 \, p_c$.
\\ ~~ \\
{\bf Fig.\ 13:}
As in Fig.\ 12, calculated with the SHASTA ($\lambda=0.4$).
\\ ~~ \\
{\bf Fig.\ 14:}
(a) $(T_c-\delta)$--isotherms in the $x-t-$plane for
$\epsilon_0 = 50, 52, 60, 100, 200, 603 \frac{1}{3}\, p_c$
(from inward to outward lying curves),
calculated with the relativistic HLLE ($\Delta x = 0.025\, R, \,
\lambda = 0.99$). Each dot
corresponds to a cell with the respective temperature.
(b) As in (a) on a larger scale, in order to show the complete
shape of the isotherm for $\epsilon_0 = 603 \frac{1}{3}\, p_c$.
(c) The lifetime of the mixed phase (i.e.\ the
intersection of the isotherms in Fig.\ 14 (a) with the $t$--axis)
as a function of $\epsilon_0$.
The solid curve (corresponding to $\epsilon_Q \geq \epsilon_0 \geq
\epsilon_H$) is calculated analytically. The diamonds are the values
as read off in Fig.\ 14 (a). The dotted line indicates the
$\epsilon_0^{3/4}$--behaviour expected for asymptotically
large $\epsilon_0$ from Bjorken's scaling solution.
\\ ~~ \\
{\bf Fig.\ 15:}
Temperature profiles for the expansion of (a--c) TN matter and (d--f) TA matter
with $\epsilon_0= 50\, p_c$ for 0, 10, 20,..., 100 time steps;
(a,d) phoenical SHASTA with original source treatment, (b,e) explicit SHASTA
with original source treatment and (c,f) with simplified source treatment.
For all cases $\Delta x = 1, \, \lambda= 0.4$.
\\ ~~ \\
{\bf Fig.\ 16:}
Temperature profiles after 0 and 100 time steps for the
expansion of (a,c) TN matter and (b,d) TA matter with
$\epsilon_0 = 50 \, p_c$, calculated with the
phoenical SHASTA with (a,b) regular antidiffusion
and (c,d) antidiffusion reduced by 20\% ($\Delta x =1,\, \lambda = 0.4$).
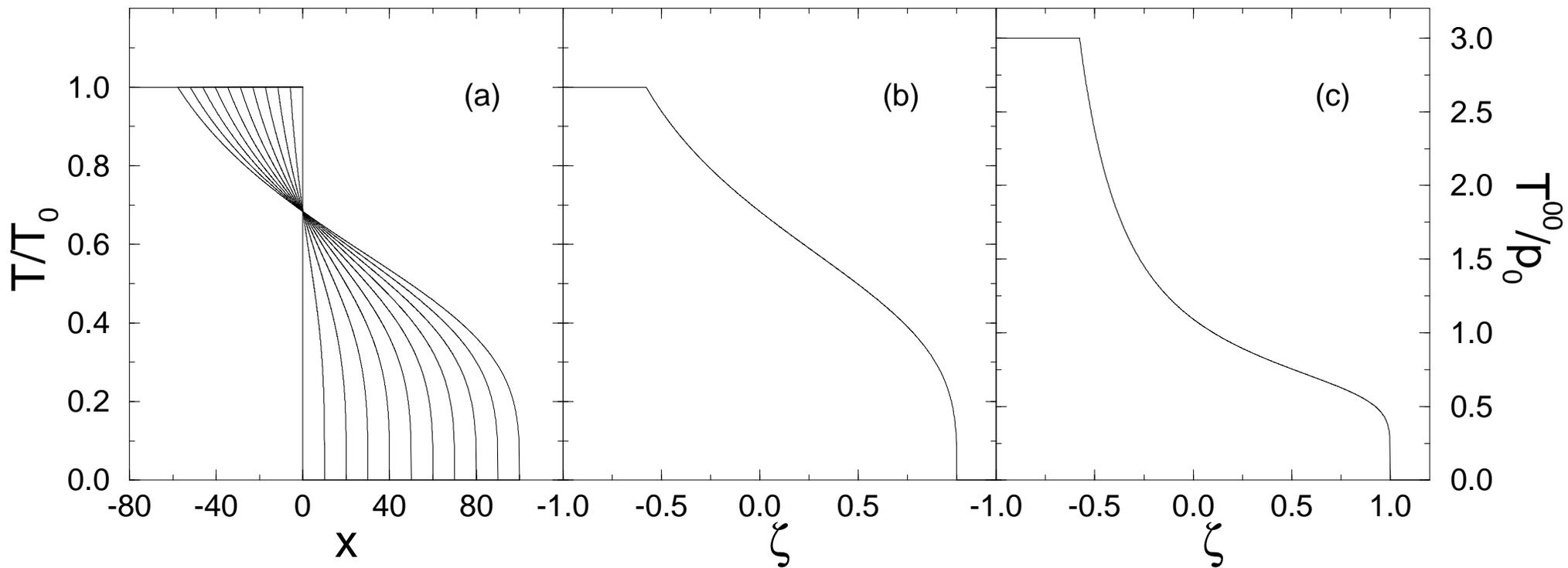
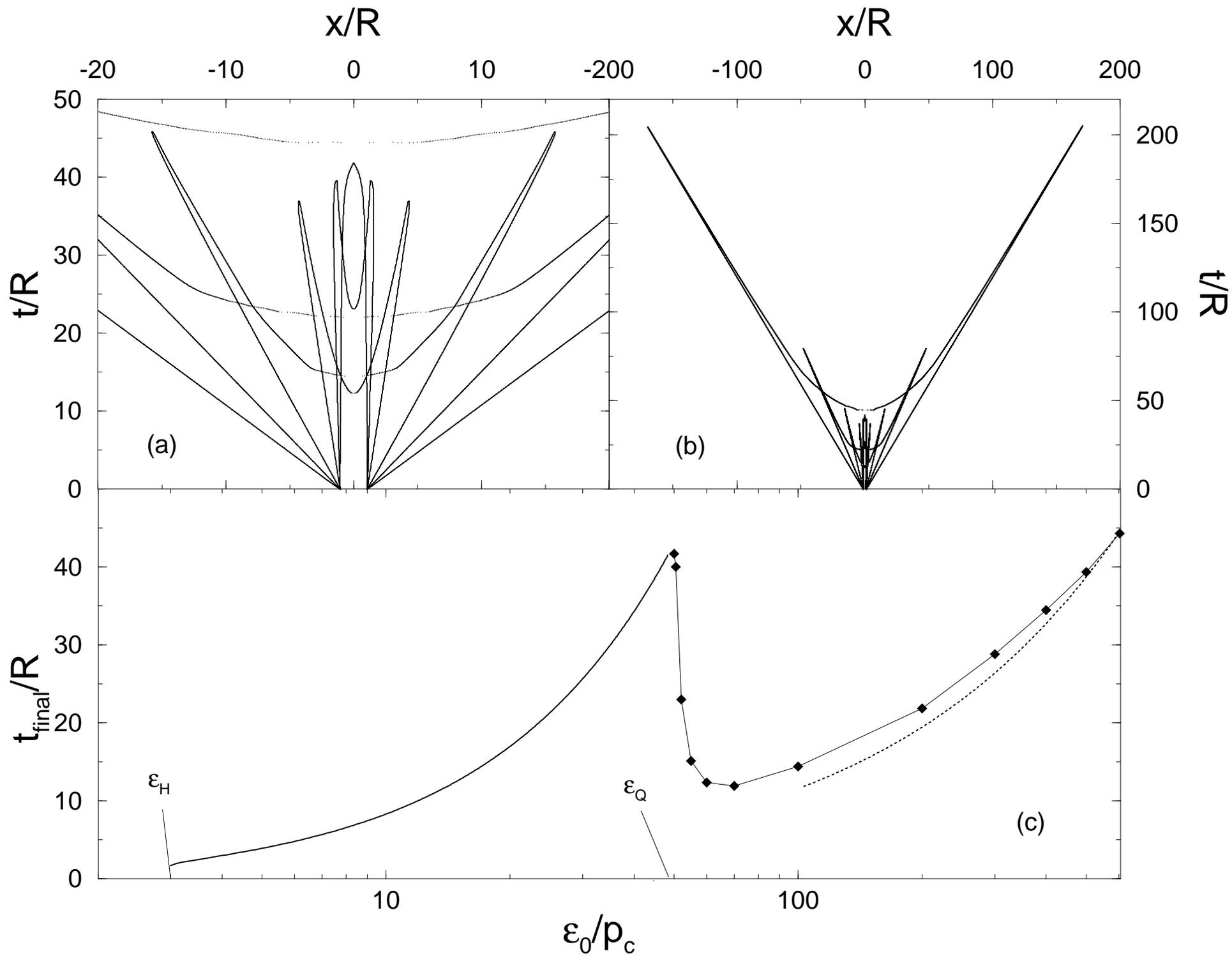
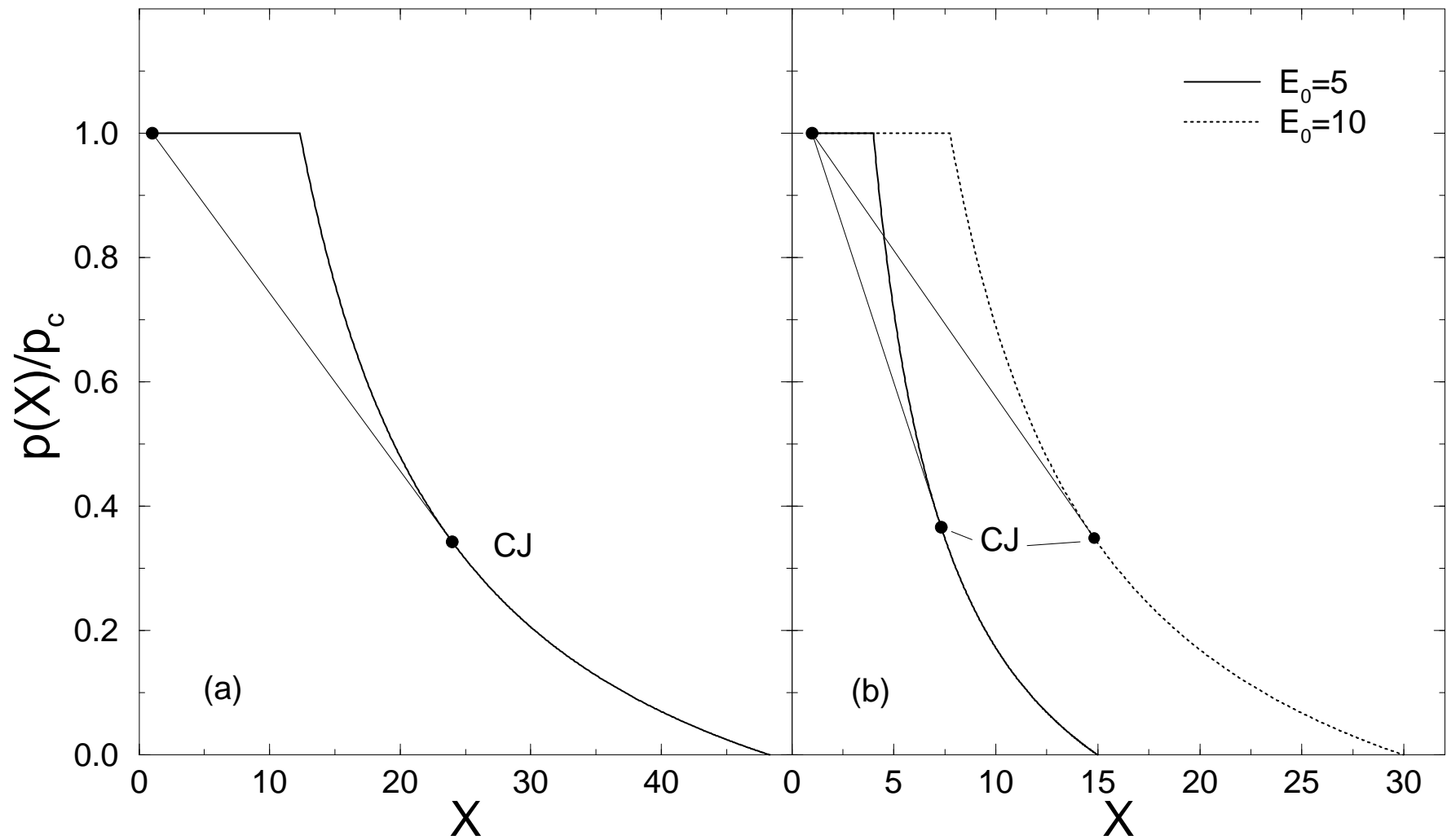
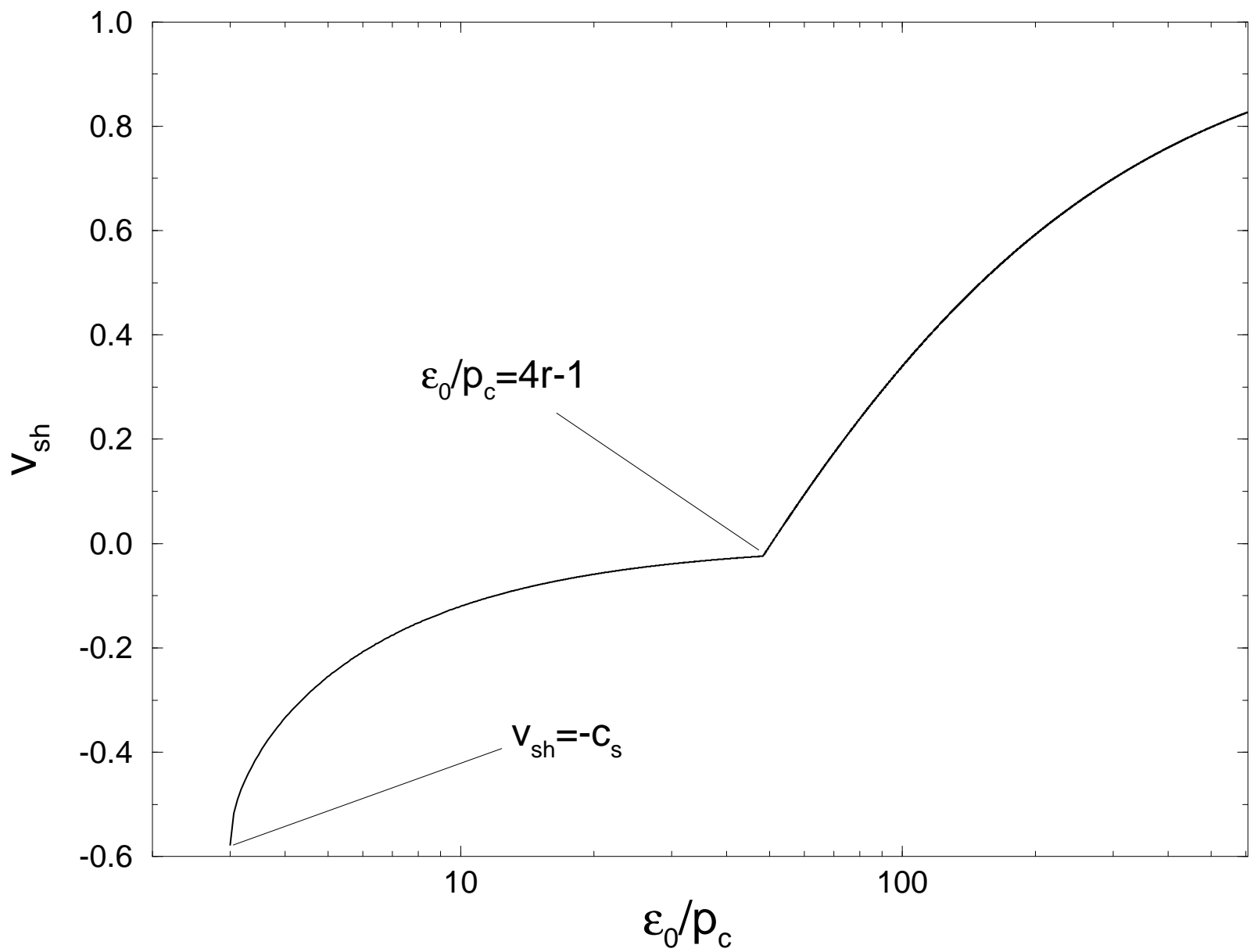
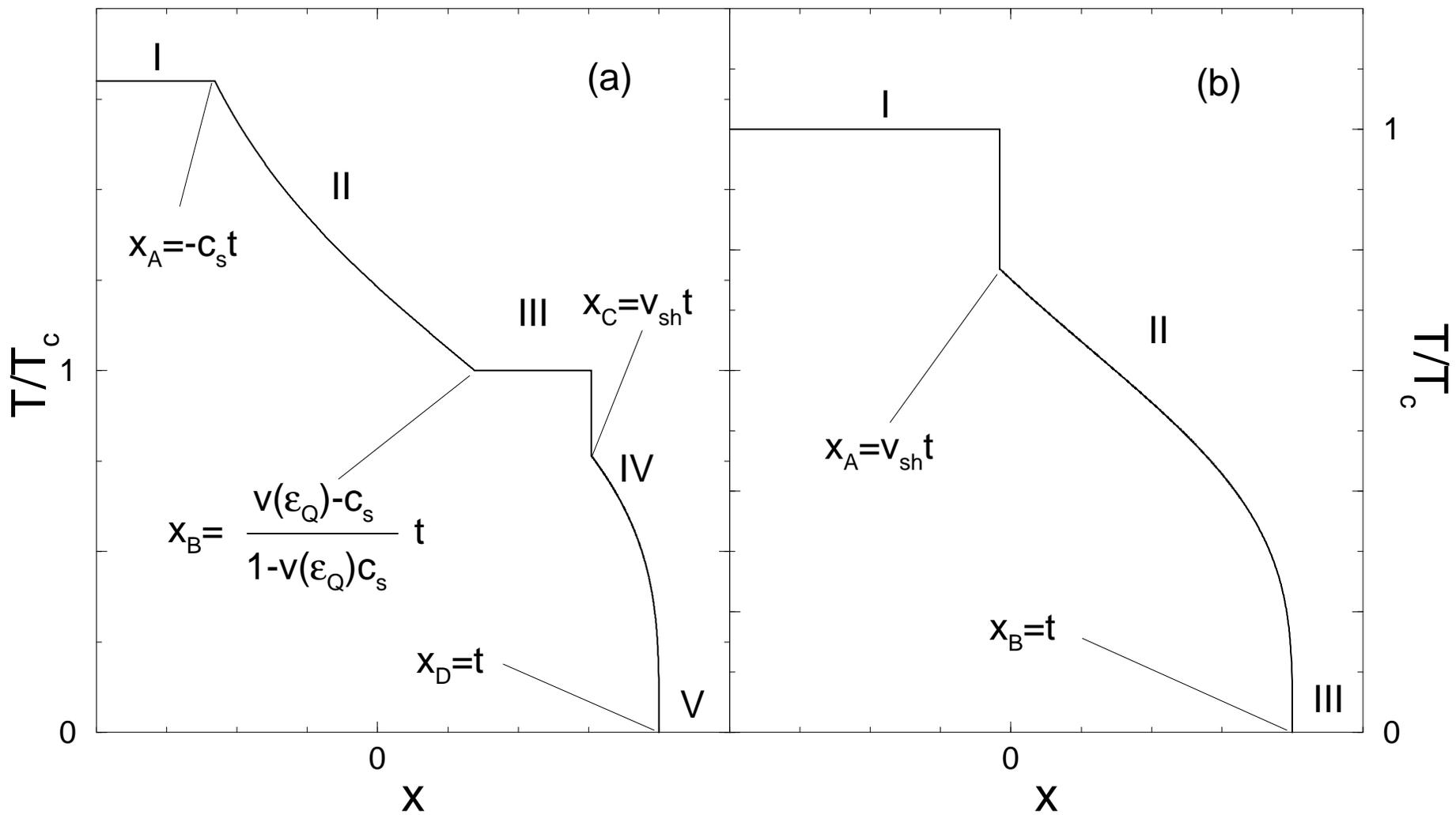
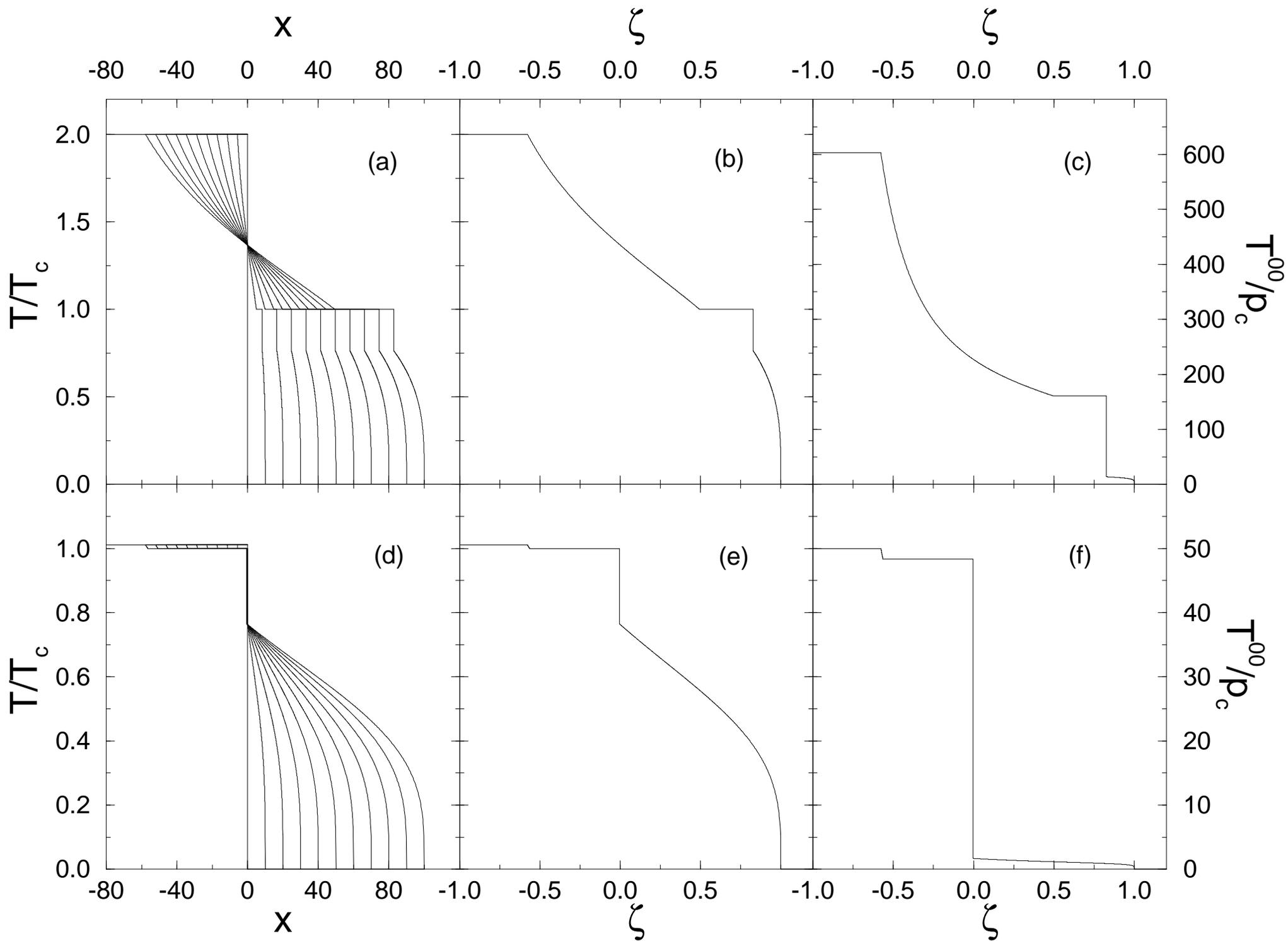
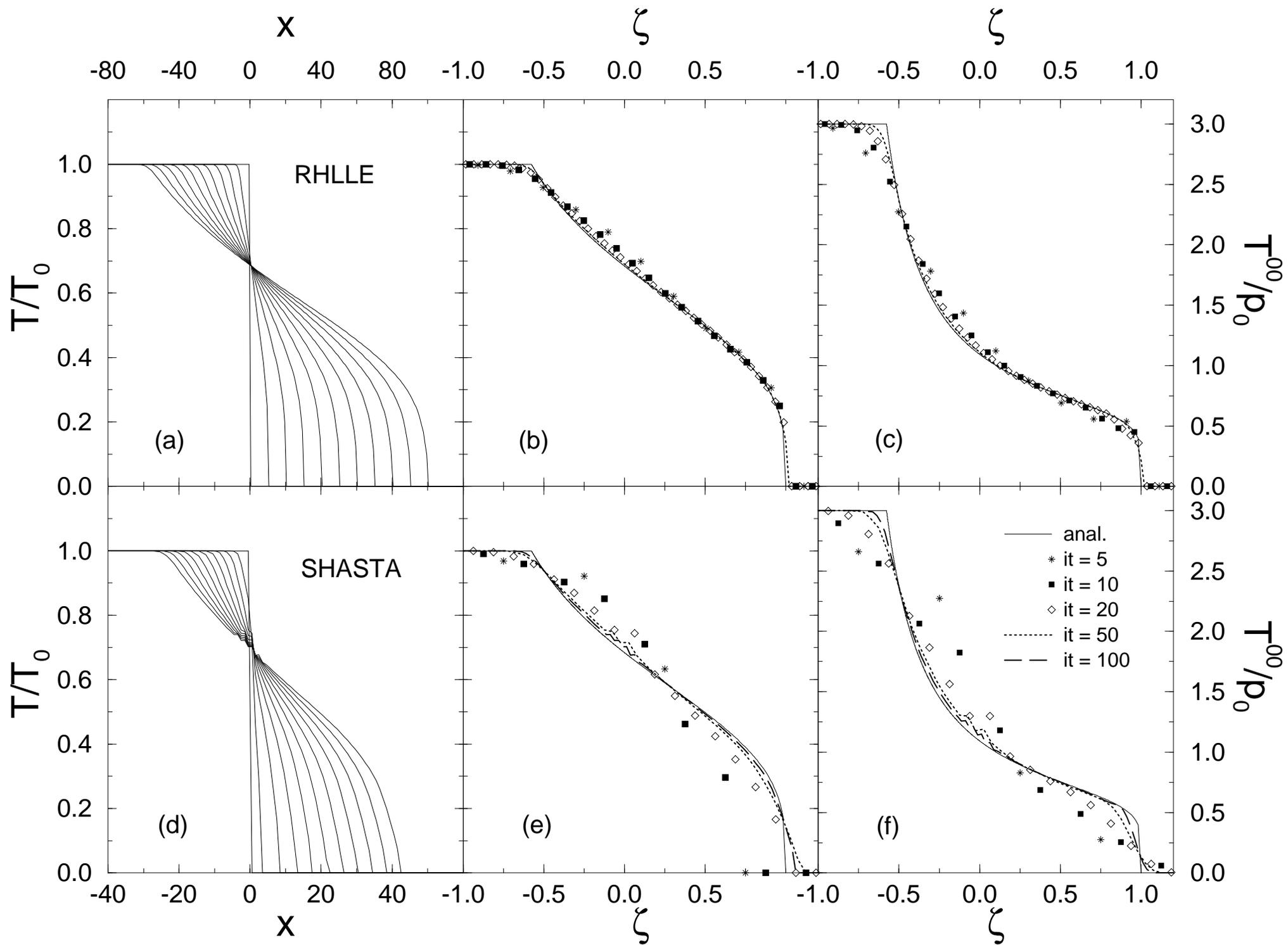
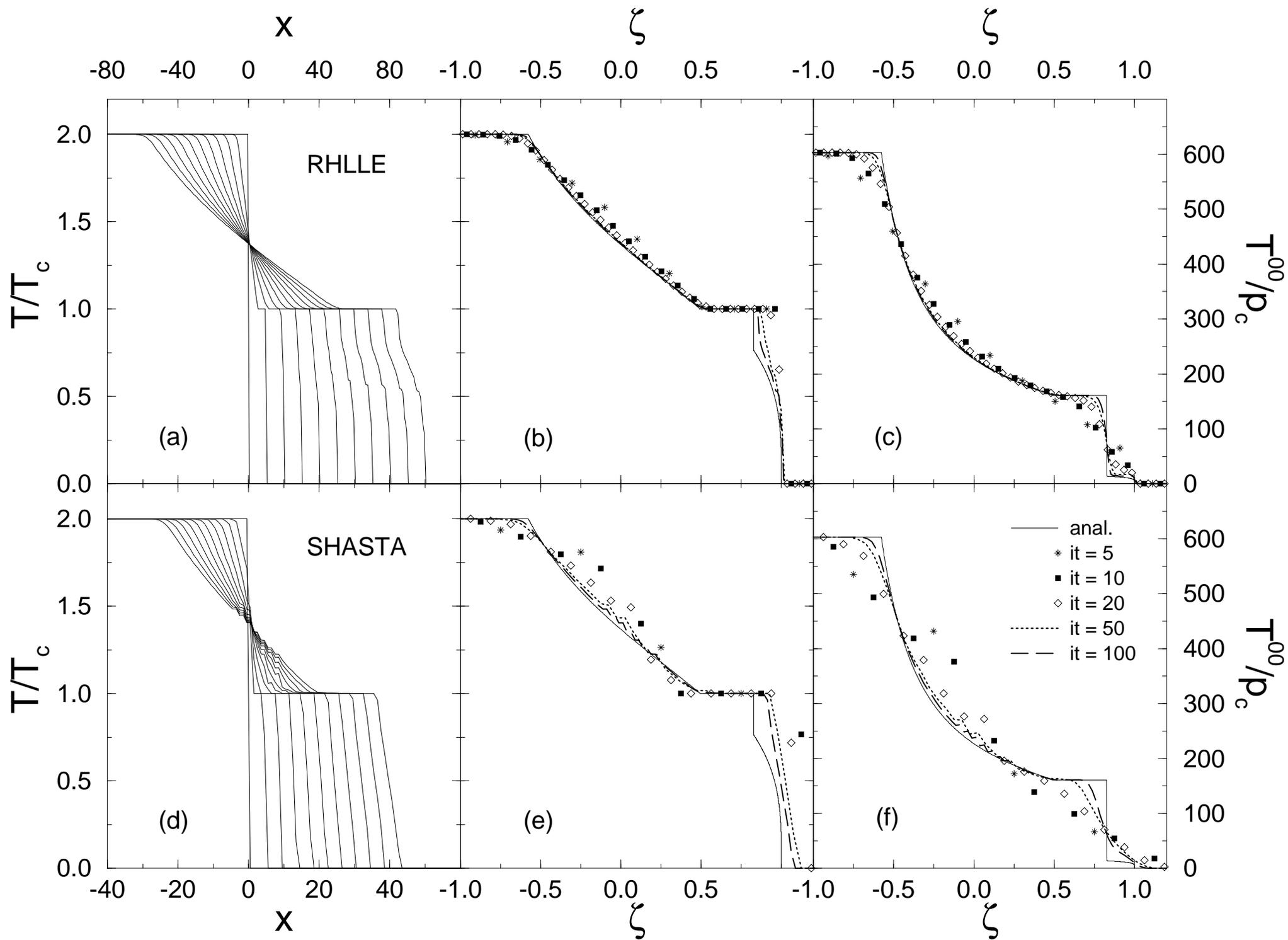
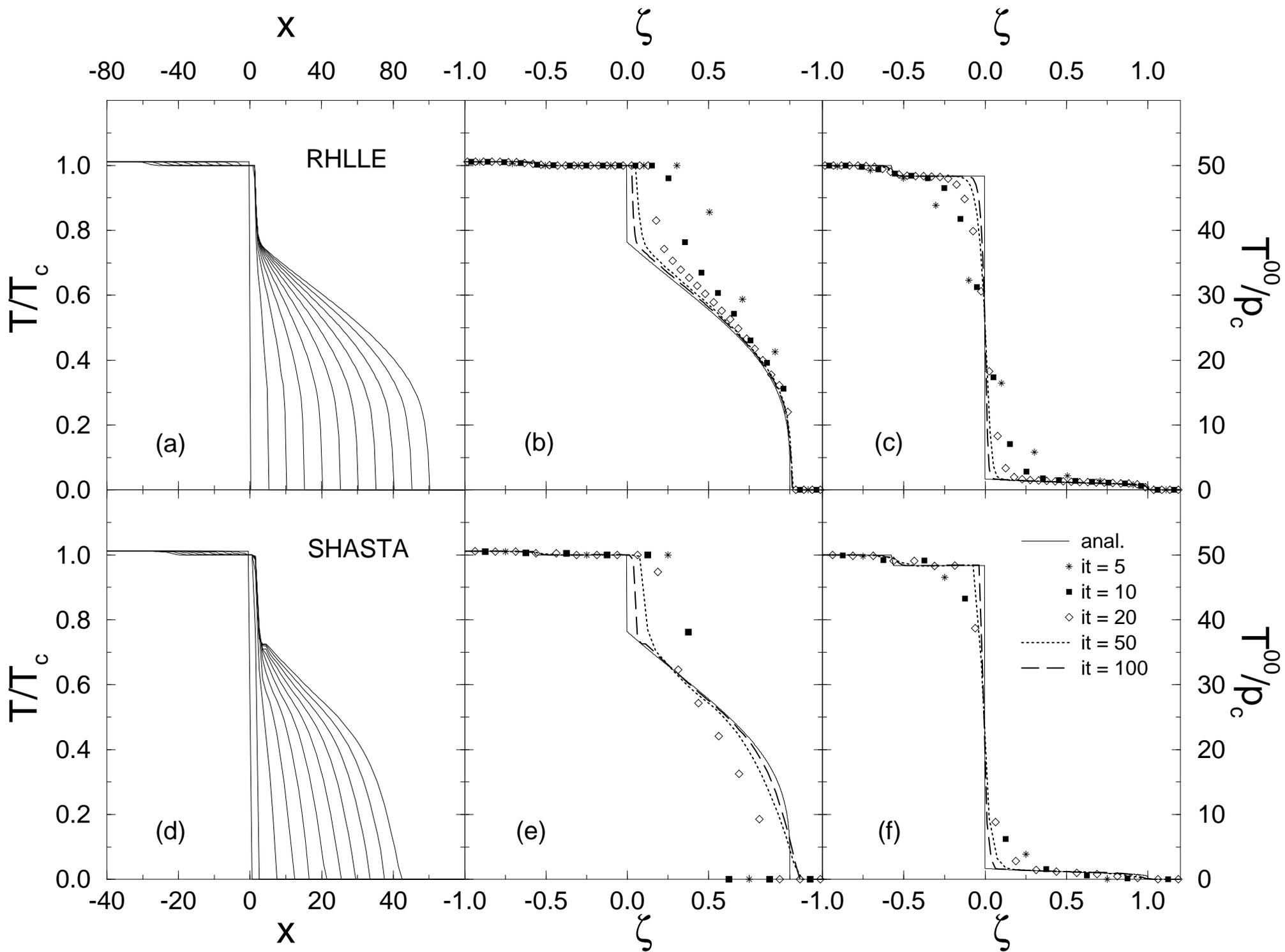
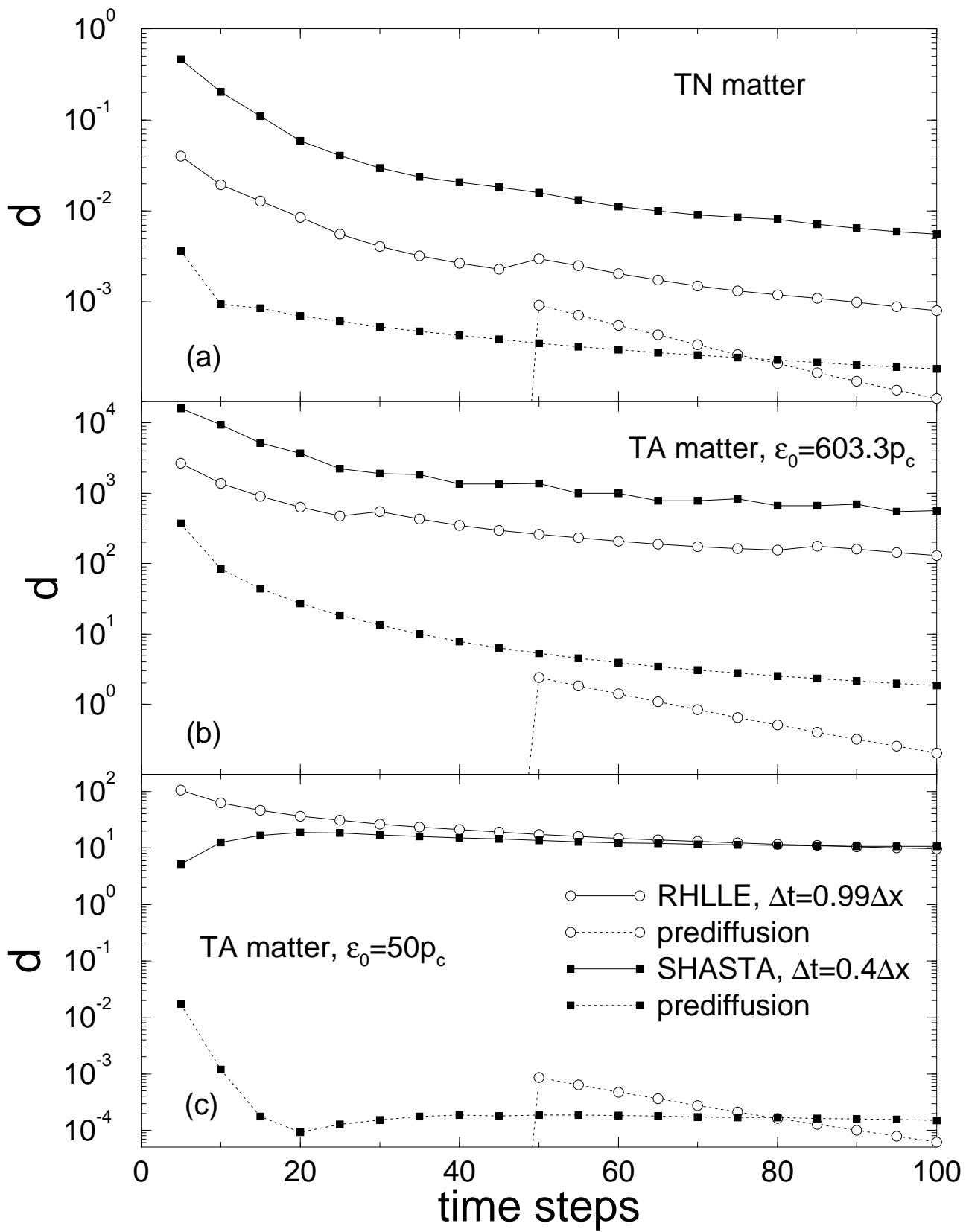
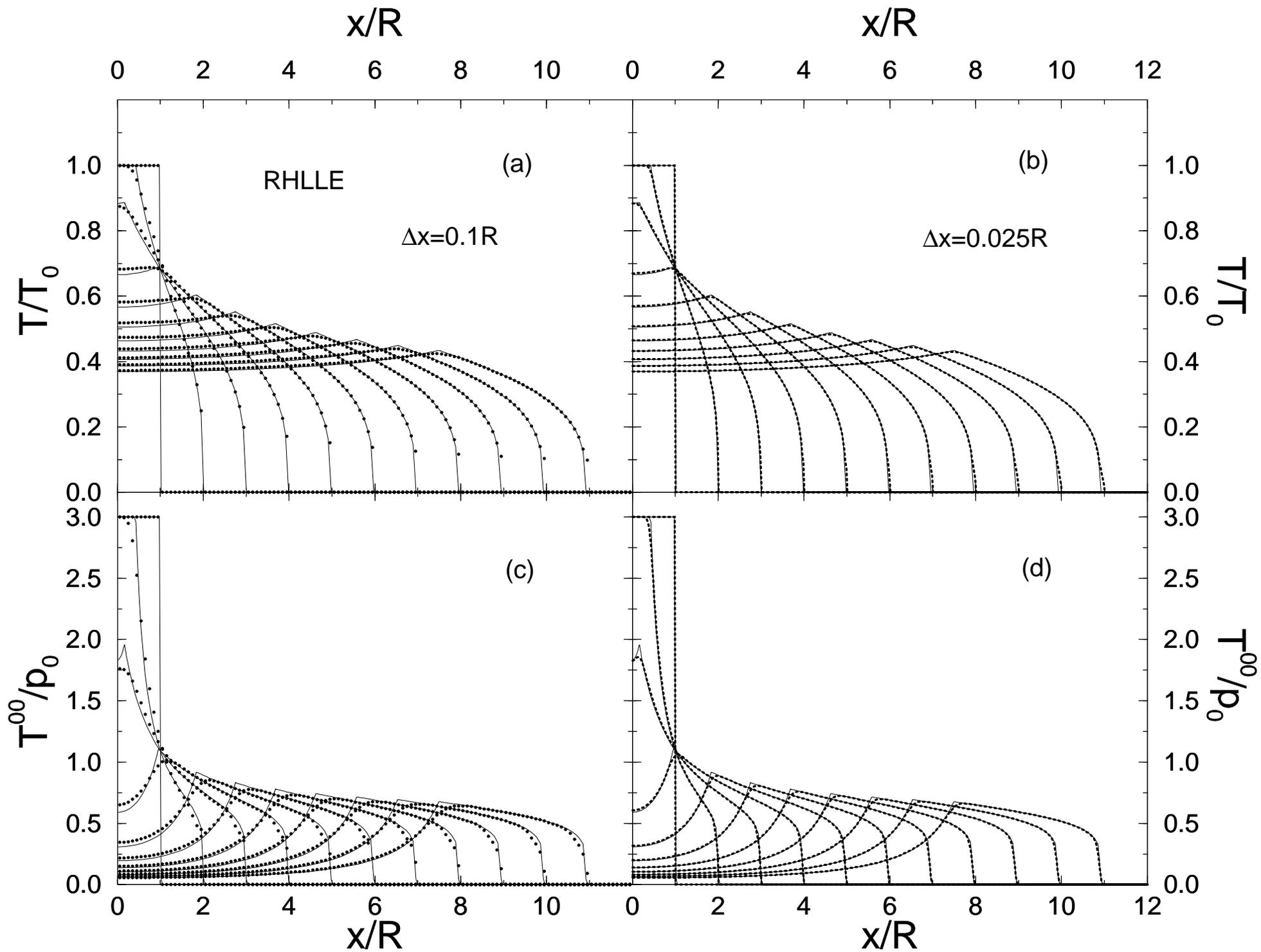
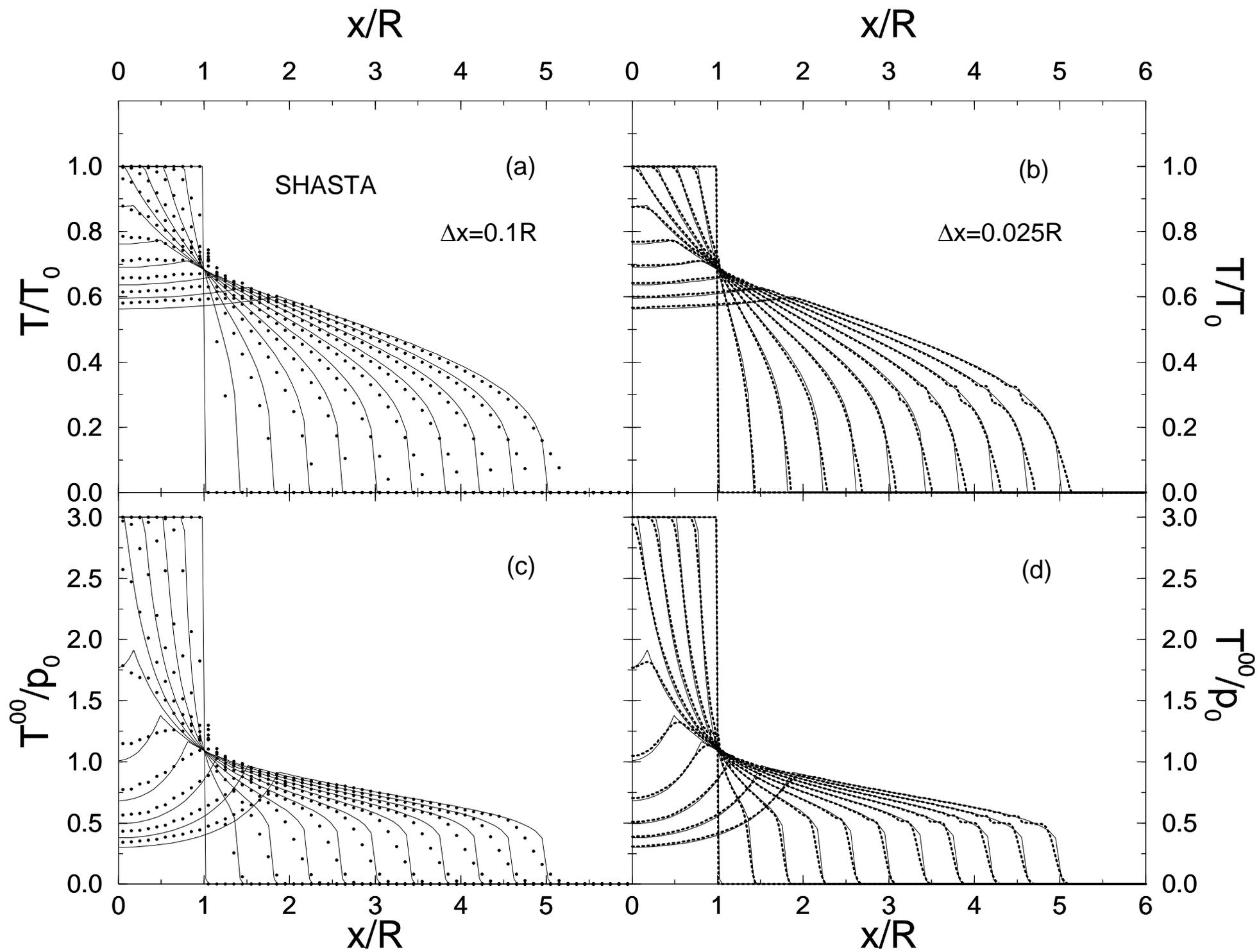
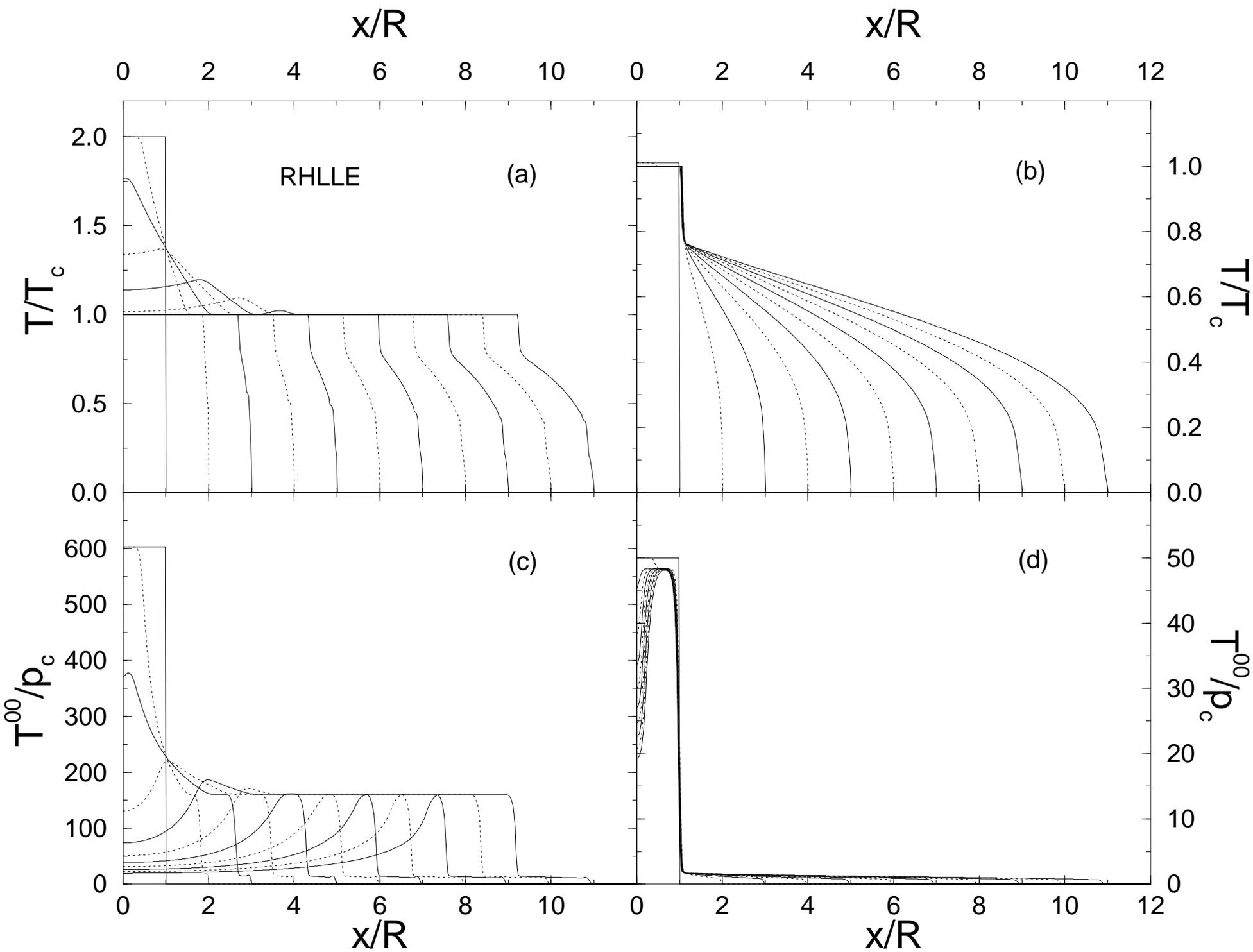
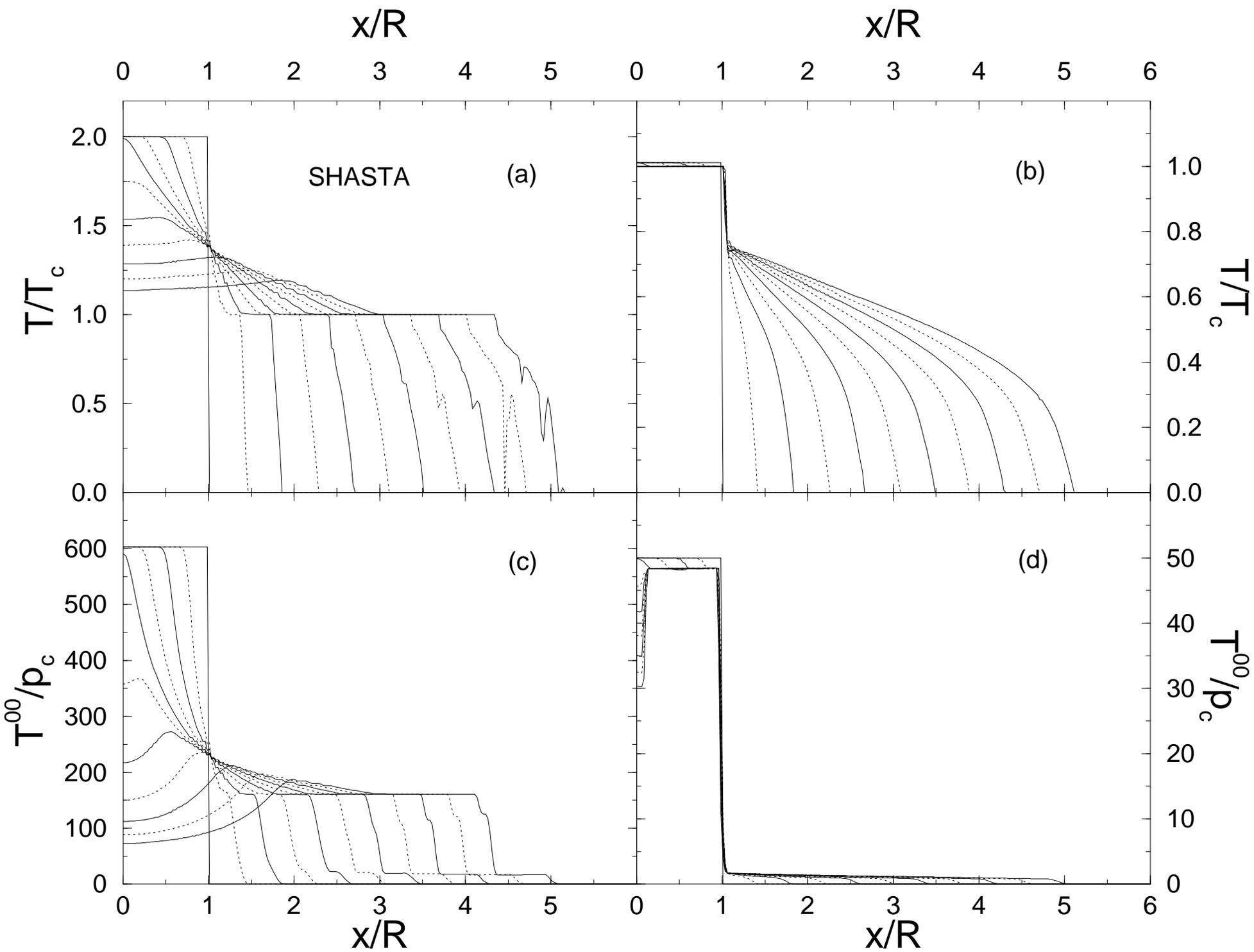
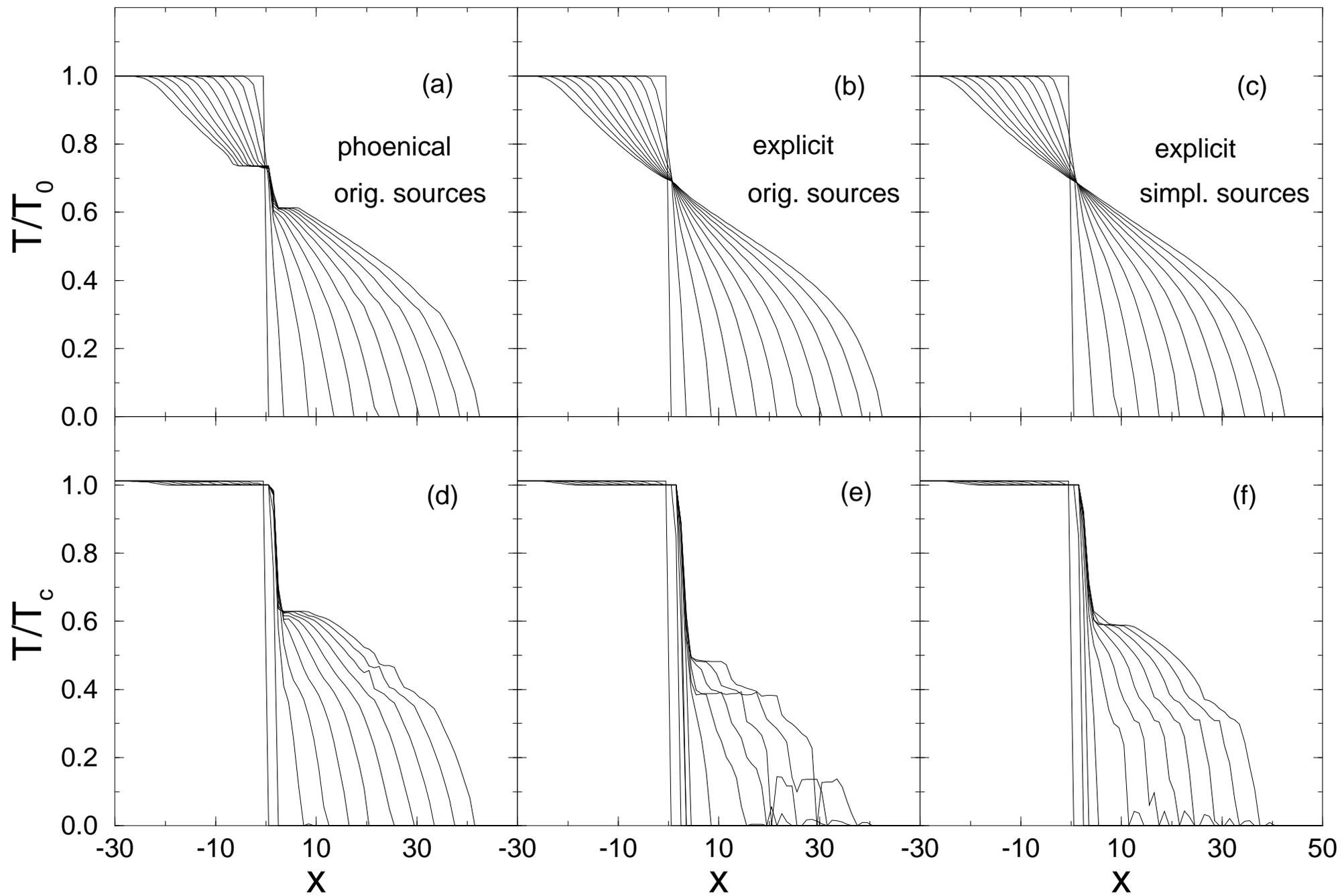
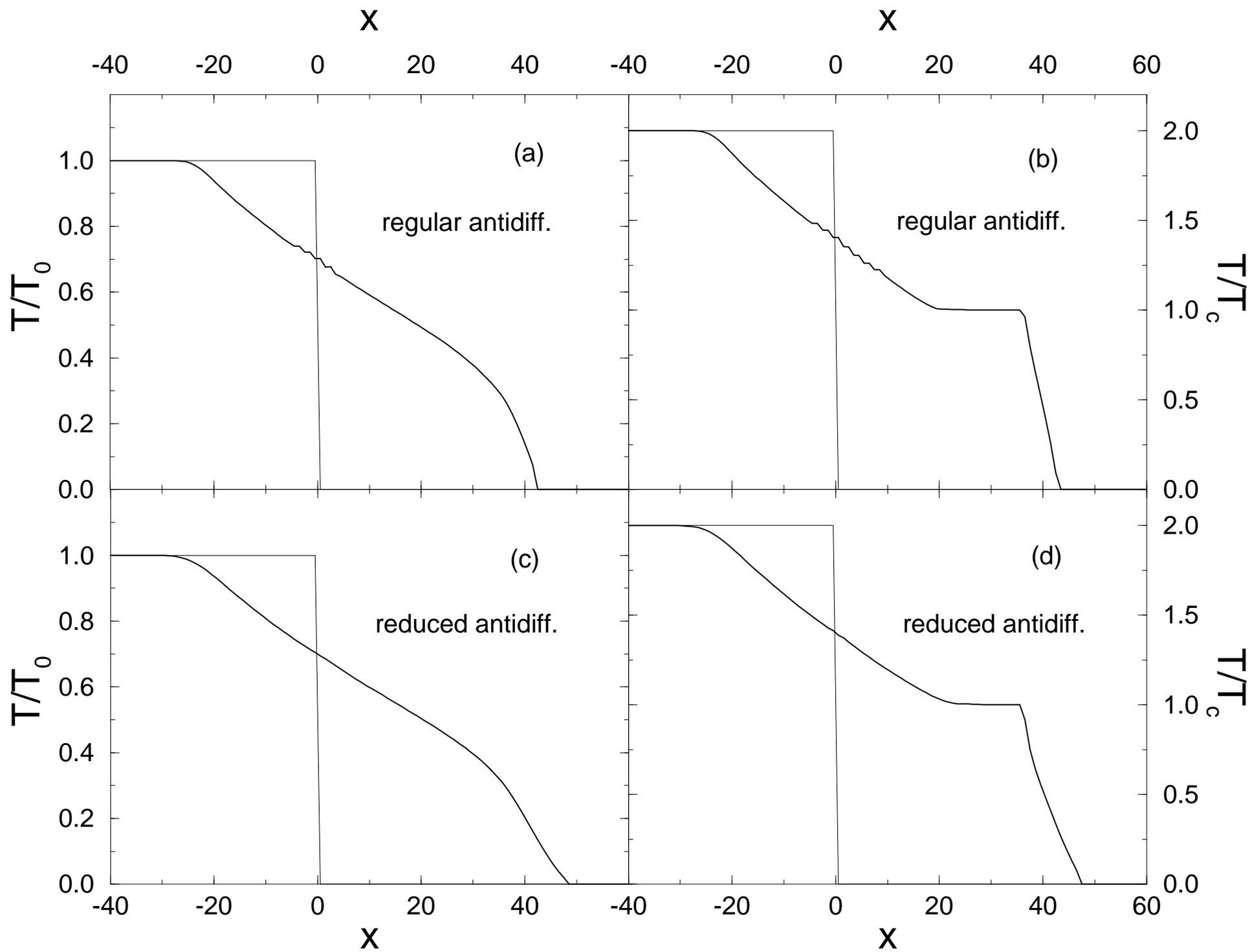
\end{document}